\documentclass[structabstract]{aa}  
\newcommand{\Ni}{{$^{56}$Ni}}
\newcommand{\Co}{$^{56}$Co}

\newcommand{\Ms}{$M_{\odot}$}
\newcommand{\grays}{$\gamma$-rays}
\newcommand{\gray}{$\gamma$-ray}
\newcommand{\arp}{Ar$^+$}
\newcommand{\hep}{He$^+$}
\newcommand{\nep}{Ne$^+$}

\newcommand{\env}{$\sim$}

\newcommand{\al}{Al$_2$O$_3$}

\newcommand{\kms}{km s$^{-1}$}
\newcommand{\cmc}{cm$^{-3}$}
\newcommand{\sidim}{Si$_2$O$_2$}
\usepackage{rotating}
\usepackage{graphicx}
\usepackage{txfonts}
\begin{document}
   \title{Molecules and dust in Cas A: I - Synthesis in the supernova phase
   and processing by the reverse shock in the clumpy remnant}

   \author{Chiara Biscaro \& Isabelle Cherchneff }

   \institute{Physik Departement, Universit{\"a}t Basel, Klingelbergstrasse 82, 4056 Basel, Switzerland }

   \date{}

 
  \abstract
   {}
   {We study the chemistry of the Type IIb supernova ejecta that led to the Cas A supernova remnant to assess the chemical type and quantity of dust that forms and evolves in the remnant phase. We later model a dense oxygen-rich ejecta knot that is crossed by the reverse shock in Cas A to study the evolution of the clump gas phase and the possibility to reform dust clusters in the post-reverse shock gas.    }
   {The chemistry is described by a chemical network that includes all possible processes efficient at high gas temperatures and densities. The formation of key bimolecular species (e.g., CO, SiO) and dust clusters of silicates, alumina, silica, metal carbides and sulphides, pure metals, and amorphous carbon is considered. A set of stiff, coupled, ordinary, differential equations is solved for the conditions pertaining to both the SN ejecta and the post-reverse shock gas.    }
   {We find that the ejecta of Type IIb SNe are unable to form large amounts of molecules and molecular clusters that are precursors to dust grains as opposed to their Type II-P counterparts because of their diffuse ejecta. The ejecta gas density needs to be increased by several orders of magnitude to allow the formation of dust clusters. We show that the chemical composition of the dust clusters that form changes drastically and gains in chemical complexity with increasing gas density. Hence, the ejecta of the Cas A supernova progenitor must have been in the form of dense clumps to account for the dust chemical composition and masses inferred from infrared observations of Cas A. As for the impact of the reverse shock on dense ejecta clumps, we show that the ejecta molecules that are destroyed by the shock reform in the post-reverse shock gas with lower abundances than those of the initial ejecta clump, except SiO. These molecules include CO, SiS and O$_2$. On the other hand, dust clusters are destroyed by the reverse shock and do not reform in the post-reverse shock gas, even for the highest gas density. These results indicate that the synthesis of dust grains out of the gas phase in the dense knots of Cas A and in other supernova remnants is unlikely.  }
   {}

   
   \authorrunning{Biscaro \& Cherchneff}
   \titlerunning{Molecules and Dust in Cas A}

   \maketitle
%

\section{Introduction}

Large amounts of cosmic dust at high redshift are inferred from the reddening of background quasars and damped Ly$\alpha$ systems (\cite{pei91,pet94}) and pose the problem of identifying the providers of dust grains in primeval galaxies and at cosmic time less than one billion years. Type II Supernovae (SNe) are strong contenders as dust factories at high-redshift because of the short evolution time-scale of their massive stellar progenitors (\env\ $10^6$ yrs). Furthermore, dust and molecules (namely carbon monoxide, CO and silicon monoxide, SiO) have been detected at infrared (IR) wavelengths in several local SNe, including SN1987A in the Large Magellanic Cloud, a few hundreds of days after the explosion (\cite{luc89,dan91,ko05}, 2006, 2009, \cite{gal12,mau12,sza13}). The dust mass derived is small and ranges from between $10^{-5}$ \Ms\ and $10^{-2}$ \Ms. These values fall short of the 1 \Ms\ of dust required per SN explosion to explain the large dust mass in primeval galaxies, as estimated by Dwek et al. (2007). 

Recently, cool dust grains have been detected with the submillimetre (submm) space telescope Herschel in several SN remnants, several decades or centuries after the SN explosion. The derived masses are larger than the masses observed in the IR at early times, and vary between \env\ 0.08 \Ms\ for the Cas A remnant (\cite{bar10}) and 0.7 \Ms\ for the young remnant SN1987A (\cite{mat11}). These large values are surprising because the physical conditions are harsh in SN remnants. The SN ejecta gas is re-processed by a reverse shock (RS) created when the mass of gas swept up by the explosion blast-wave exceeds the ejecta mass (\cite{chev77}). The RS travels inwards and therefore induces the partial destruction of the molecules and dust synthesised and expelled in the SN ejecta phase. 

Owing to its proximity (3.4 kpc) and angular size (\env~5') (\cite{ree95}), the SN remnant Cas A has been extensively studied observationally. Cas A results from the explosion of a SN of type IIb 330 years ago. The progenitor was a massive supergiant star having lost most of its hydrogen envelope and with a ZAMS mass comprises between 15 and 25 \Ms\ (\cite{krau08}). The explosion blast wave travelled in a diffuse medium, leading to a low density ejecta compared to regular Type II SNe (\cite{fil88}). The Cas A ejecta is now crossed by the RS and warm dust has been detected with the IR space telescope Spitzer in a mass range $0.02 - 0.054$ \Ms\ at the RS position (\cite{rho08}). Cool dust interior to the RS position has also been detected with Herschel (\cite{bar10}), with an estimated mass of \env\ 0.075 \Ms. Of prime importance is the detection of molecules in the remnant. Warm carbon monoxide, CO, has been observed at the RS position with Spitzer (\cite{rho09}, 2012) and in one dense clump with Herschel (\cite{wal13}). In the young remnant SN1987A, cool CO has been recently observed with ALMA along with the partial spectroscopic detection of SiO (\cite{kam13}). This detection points to chemical species, CO and SiO, formed in the ejecta after the SN explosion. As in other evolved circumstellar environments, molecules appear to be strongly coupled to the formation of dust grains in SNe (\cite{cher10b}). Once more, the presence of dust and molecules in Cas A and other remnants confirms this inherent aspect of the synthesis of cosmic dust and indicates a strong chemical reprocessing of the material ejected during the SN phase by the RS. 

\begin{table*}[]
\caption{Thermal and non-thermal processes and chemical reaction types included in the chemical model for both the Type IIb SN ejecta and the post-reverse shock gas. }
\label{tab1}   
\centering
\begin{tabular}{l rl l l l}
\hline \hline
\multicolumn{5}{c}{Reaction description} & Gas regime/location \\
\hline 
\multicolumn{6}{c}{THERMAL } \\
\hline
Unimolecular & AB &$\longrightarrow$& A + B & Thermal decomposition & Very high temperature\\
\hline
Bimolecular & AB+ C &$\longrightarrow$& BC + A & Neutral exchange & High temperature \\
& A + B &$\longrightarrow$&  AB + h$\nu$ & Radiative association & T independent \\
& AB + M &$\longrightarrow$& A + B + M & Collision dissociation & High density\\
& AB$^+$ + C &$\longrightarrow$&  BC$^+$ + A & Ion$-$Molecule & T independent \\
& AB$^+$ + C &$\longrightarrow$&  AB + C$^+$ & Charge exchange & T independent\\
& A $^+$ + e$^-$ &$\longrightarrow$& A + h$\nu$&Radiative recombination&  T independent \\
& AB $^+$ + e$^-$ &$\longrightarrow$& A + B& Dissociative recombination&  T independent \\
\hline
Termolecular & A + B + M &$\longrightarrow$&  AB + M & Three-body association & High density\\

\hline
\multicolumn{6}{c}{NON-THERMAL } \\
\hline
& A + CE &$\longrightarrow$&  A$^+$ + e$^-$ +CE& Ionisation by Compton e$^-$  & entire SN ejecta \\
& AB + CE &$\longrightarrow$&  A + B + CE& Dissociation by Compton e$^-$  & entire SN ejecta \\
& AB + h$\nu$ &$\longrightarrow$&  A + B & Photo-dissociation & post-reverse shock gas \\
& A + h$\nu$ &$\longrightarrow$&  A$^+$ + e$^-$& Photo-ionisation & post-reverse shock gas \\
\hline
\end{tabular}
\end{table*}
Theoretical studies on both the formation of dust in SNe and the reprocessing by the RS have been carried out. Models of dust production in SN ejecta lead to contradictory results on the dust quantity produced after the SN explosion. Early studies based on classical nucleation theory (CNT) indicate large dust masses (\cite{koz89, tod01, noz03}) that cannot be reconciled with IR observations of SNe. Models based on chemical kinetics and considering the coupling between molecules and dust grains result in lower amount of dust formed in the nebular phase (\cite{cher08, cher10a}). More recently, Sarangi \& Cherchneff (2013, hereafter SC13) show that the molecular clusters precursors to dust (hereafter, dust clusters) gradually grow from low to high masses on a time-span of a few years after outburst and provide a genuine explanation to the discrepancy on dust mass derived from IR data of SNe and submm data of SN remnants. The reprocessing of dust by the RS in SN remnants has been modelled assuming pre-shock dust distributions derived from CNT and in the context of a homogeneous SN ejecta (\cite{noz10, sil12}). None of these studies addresses the chemistry of the RS, the survival of molecules and the possibility to reform dust after the passage of the RS. Therefore, whether SNe and SN remnants are dust makers and/or destroyers is still unclear on the basis of both available theoretical predictions and observational data. 

In order to advance our understanding of the net dust budget produced by SNe, we carry out a global study of the synthesis of molecules and dust in SN ejecta and their reprocessing by the RS in the remnant phase, focusing on Cas A. In this paper (pap. I),  we study the formation of molecules and dust clusters in the Type IIb SN that led to Cas A, and the later processing of the produced gas-phase material by the RS. In a forthcoming publication (pap. II), we will address the condensation of dust clusters, and derive grain size distributions for the Type IIb SN. We will then study the dust reprocessing in clumps for various RS models. Presently, we model the non-equilibrium chemistry of the SN ejecta and that of the post-RS gas, assuming that the processed ejecta material resides in dense, fast-moving knots, as observed by Fesen et al. (2001, 2006) and Docenko \& Sunyaev (2010, hereafter DS10). In Section 2, we present the chemistry considered in this study for both the SN type IIb and the dense, shocked, remnant knots. The physical models used for the SN ejecta and the knot shocked by the RS are described in Section 3. We present the results in Section 4 and discuss our findings in section 5.  

\section{Chemical model}
\label{chem}

\begin{table*}[]
\caption{Chemical species and dust clusters included in the chemical model of both the Type IIb SN ejecta and the post-reverse shock gas.}
\label{tab2}   
\centering
\begin{tabular}{l l l l l l l l l l l  }
\hline \hline
\multicolumn{10}{c}{{\bf Atoms} }\\

\hline

O & Si & S & C & N & Mg & Al& Fe & He & Ne & Ar \\
\hline
\multicolumn{10}{c}{{\bf Ions}} \\
\hline

O$^+$ & Si$^+$ & S$^+$ & C$^+$ & N$^+$ & Mg$^+$ & Al$^+$& Fe$^+$ & He$^+$ & Ne$^+$ & Ar$^+$ \\ 
C$_2^+$&SiO$^+$ & CO$^+$ & O$_2^+$ & SO$^+$  & & & & & \\
\hline
\multicolumn{10}{c}{{\bf Molecules}} \\
\hline

O$_2$ & CO & SiO & SO & NO& AlO & FeO & MgO & CO$_2$ & CN & CS  \\ 
SiS & SiC &  FeS & MgS & S$_2$ & N$_2$ & & \\
\hline
\multicolumn{10}{c}{{\bf Dust clusters}} \\
\hline

C$_2$ & C$_3$ & C$_4$& C$_5$ & C$_6$ & C$_7$ & C$_8$ & C$_9$ & C$_{10}$ & &\\
Si$_2$ & Si$_3$ & Si$_4$& Mg$_2$ & Mg$_3$ & Mg$_4$& Fe$_2$ & Fe$_3$ & Fe$_4$ & Al$_2$ &\\
Si$_2$C$_2$ & Mg$_2$S$_2$ & Fe$_2$S$_2$ & Fe$_3$S$_3$ & Fe$_4$S$_4$& &&& &\\
Si$_2$O$_2$& Si$_3$O$_3$& Si$_4$O$_4$&Si$_5$O$_5$ & SiO$_2$ & Si$_2$O$_3$&  Si$_3$O$_4$ & Si$_4$O$_5$ & & \\ 
MgSi$_2$O$_3$ & MgSi$_2$O$_4$ & Mg$_2$Si$_2$O$_4$ & Mg$_2$Si$_2$O$_5$ & Mg$_2$Si$_2$O$_6$ & Mg$_3$Si$_2$O$_6$ & Mg$_3$Si$_2$O$_7$ & Mg$_4$Si$_2$O$_7$ & Mg$_4$Si$_2$O$_8$ &\\
Mg$_2$O$_2$ & Mg$_3$O$_3$ & Mg$_4$O$_4$ &Fe$_2$O$_2$ & Fe$_3$O$_3$ & Fe$_4$O$_4$ & Al$_2$O$_2$&  AlO$_2$ &Al$_2$O$_3$ & Al$_4$O$_6$ \\

\hline
\end{tabular}
\end{table*}

The synthesis of dust in SN environments stems from the chemical ability of the elements produced via nucleosynthesis during the stellar evolution of the massive progenitor and the SN explosion to assemble into molecules and dust molecular clusters. SN ejecta are characterised by extreme physical conditions including high gas temperature and velocity in the back of the explosion blast wave, radioactivity induced by the decay of \Ni, and \grays\ and ultraviolet (UV) fields. For both environments studied in this paper, we use similar processes as in the study of dust formation in primeval and Type II-P SNe by Cherchneff \& Dwek (2009, 2010) and SC13, respectively. A chemical kinetic approach is based on a chemical reaction network that includes the thermal and non-thermal processes summarised in Table \ref{tab1}. Thermal processes include unimolecular, bimolecular, and termolecular reactions. Because of the high gas temperatures present a few months after the SN explosion, neutral exchange reactions, which often have activation energy barriers reflecting the energy required to break and re-arrange chemical bonds, are prevalent formation processes in the build-up of chemical complexity and the formation of dust clusters (\cite{cher09,cher10a}, SC13). 

In addition to these reactions, we consider non-thermal processes that include the destruction of molecules and ionisation of atoms by energetic Compton electrons created when the \gray\ photons generated by the decay of \Ni\ and \Co\ degrade via collisions with electrons. Rates are estimated following Cherchneff \& Dwek (2009) and we assume an average \gray\ optical depth for the ejecta with a 19~\Ms\ stellar progenitor (as shown by SC13, the use of a \gray\ optical depth for each ejecta zone as calculated by Kozma \& Fransson (1992) for a 20~\Ms\ progenitor has little impact on the trends and results derived for an average \gray\ optical depth). Photo-ionisation and -dissociation by UV radiation are considered in the chemistry of the shocked clump. 

The various atoms, molecules, and ions considered to participate in the SN ejecta chemistry are summarised in Table \ref{tab2}. Small molecular clusters form according to the processes described in SC13, where the description of the growth pathways of small silicate clusters, namely forsterite dimers (Mg$_2$SiO$_4$)$_2$ and enstatite dimers (MgSiO$_3$)$_2$, is based on the work by Goumans \& Bromley (2012). This study indicates possible chemical routes to the formation of the silicate dimers, which involve the formation of the SiO dimer ring, (SiO)$_2$,  and its growth to Si$_2$O$_3$ through the reaction with O$_2$ and SO. The subsequent pathway involves the addition of a magnesium atom to the Si$_2$O$_3$ structure. The later growth of clusters is described by one O-addition step followed by one Mg inclusion as a recurrent growth scenario. We have also considered a new pathway for the formation of the alumina dimer, (Al$_2$O$_3$)$_2$, that was not considered in SC13. Because the structure of the molecule \al\ is akin to that of Si$_2$O$_3$, we have considered a similar growth route starting from the formation of AlO, and followed by its dimerisation to Al$_2$O$_2$ and the formation of \al\ via oxidation. The alumina dimers are then formed from the recombination of two \al\ molecules. Oxidising agents for the formation of silicates and alumina include atomic O, O$_2$, AlO and SO, but reactions with O$_2$ and SO are prevalent for silicate clusters while reaction with AlO dominates in the case of alumina. 

The chemical modelling of the Type IIb SN that led to Cas A includes two temperature regimes and chemical networks, because the initial gas temperature and density quickly drop with time from 100 days to 3000 days (\env~8 years) post-explosion (see next Section and Table \ref{tab4}). These values are lower than in the case of Type II-P SNe because the ejecta quickly expands at high velocities owing to the lack of a progenitor circumstellar wind that would slow down the explosion blast wave through interaction. We thus consider a high temperature reaction network from day 100 (\env\ 7000 K) to day 1000 (300K) and a low temperature network from day 1001 until day 3000 (\env\ 60 K), for which the reaction rates of relevant reactions have been adjusted according to available, low-temperature data. 

The chemical kinetic description is applied to the gas parameters that characterise the SN ejecta and shock models (see Section \S\ \ref{phys}). The variation of the number density of the chemical species $i$ with time is described by the following rate equation

\begin{equation}
\label{eq1}
\frac{dn_i}{dt} = P_i - L_i = {\sum}_j k_{ij}n_jn_i - {\sum}_k k_{ik}n_in_k
\end{equation}
where $P_i$ and $L_i$ are the total chemical production and loss processes for species $i$, $n_i$ is the number density for species $i$, and $k_{ij}$ the rate for the reaction of $i$ with $j$. This rate is expressed in Arrhenius form as 

\begin{equation}
\label{eq2}
k_{ij}(T)=A_{ij} \times \left( \frac{T}{300} \right)^{\nu} \times exp(-E_{ij}/T)
\end{equation}
where T is the gas temperature in Kelvin, $\nu$ the temperature dependence exponent, and $E_{ij}$ the activation energy barrier in Kelvin. The coefficient A$_{ij}$ has the units of s$^{-1}$ for unimolecular processes, cm$^3$ s$^{-1}$ for bimolecular reactions, and cm$^6$ s$^{-1}$ for termolecular reactions.

The system includes 93 species, 412 chemical reactions, and a set of 93 stiff, coupled, ordinary differential equations is solved using a Gear method (\cite{hind83}).
 

\section{Physical models}
\label{phys}
\begin{table*}
\label{tab3} 
\caption{Gas mean molecular weight $\mu_{gas}$, C/O ratio, and initial elemental mass yields as a function of ejecta zone for the 19 \Ms\ progenitor of Rauscher et al. (2002). The total yield is the elemental mass yield for the total ejected material.}
\centering
\begin{tabular}{l l l l l   l l l l l l l l l}
\hline \hline
\multicolumn{1}{c}{Zone} & \multicolumn{1}{c}{$\mu_{gas}$}& \multicolumn{1}{c}{C/O}& \multicolumn{1}{c}{He} & \multicolumn{1}{c}{C}& \multicolumn{1}{c}{O} &\multicolumn{1}{c}{Ne} & \multicolumn{1}{c}{Mg}&\multicolumn{1}{c}{Al}& \multicolumn{1}{c}{Si}& \multicolumn{1}{c}{S }&\multicolumn{1}{c}{Ar}& \multicolumn{1}{c}{Fe} & \multicolumn{1}{c}{Ni} \\

\hline
1a & 42.3 &0.08 & 0 & 1.0(-7) & 1.3(-6) & 6.4(-7) & 1.8(-5) & 2.7(-5) &3.9(-2) & 2.3(-2) & 4.7(-3)& 5.8(-3) & 1.1(-1)\\
1b & 22.5 & 1(-3)& 0 & 1.1(-4) & 1.2(-1) & 1.0(-4) & 8.7(-4) & 2.2(-4) &9.8(-2) & 5.6(-2) & 1.5(-2) & 2.7(-3) & 6.1(-4) \\
2  & 17.0 & 0.049& 0 & 5.7(-2) & 1.1 & 3.3(-1) & 8.4(-2) & 9.1(-3) & 1.5(-2) & 1.3(-3) & 1.2(-4) & 7.5(-4) & 2.2(-4)  \\
3a & 15.1 &0.27 & 0 & 2.9(-2) & 1.1(-1) & 2.8(-3) & 2.0(-3) & 1.6(-5)  & 1.1(-4) & 3.2(-5) & 8.9(-6) & 7.8(-5) & 2.8(-5) \\
3b & 10.3 &0.48 & 7.4(-2) & 1.2(-1) & 2.5(-1) & 1.3(-2) & 5.2(-3) & 5.1(-5)  & 3.6(-4) & 1.3(-4) & 3.3(-5) & 3.6(-4) & 2.8(-5) \\
4a & 4.1 &3.0 & 7.2(-1) & 1.2(-2) & 4.2(-3) & 1.0(-2) & 4.8(-4) & 5.8(-5)  &  5.4(-4) &3.1(-4) & 6.8(-5) & 9.5(-4) & 5.2(-5) \\ 
4b & 4.0&1.4 & 3.3(-1) & 1.2(-4) & 8.8(-5) & 5.4(-4) & 2.1(-4) & 3.0(-5)&2.4(-4) & 1.4(-4) & 3.2(-5) & 4.3(-4) & 2.4(-5)\\
\hline
\multicolumn{3}{l}{Total Yield} & 1.124& 0.218&1.584  &0.356 &9.3(-2) &9.5(-3) &0.153 & 8.1(-2)&2.0(-2) & 1.1(-2)&  0.111 \\
\hline
\end{tabular}
\tablefoot{Mass yields are in \Ms\ and the gas mean molecular weight $\mu_{gas}$ is in g \cmc. Nitrogen, N, is only present in Zone 4b with a yield of $4.2\times 10^{-3}$.}
\end{table*}

We study the chemistry of two distinct environments: 1) the ejecta of the SN Type IIb that led to the remnant Cas A, and 2) one oxygen-rich knot in Cas A that is crossed by the reverse shock. The aim is to derive a comprehensive picture of the production and processing of molecules and dust in Cas A and to assess whether the chemical species synthesised in SNe survive to reverse shock processing in the remnant phase. First, we present our physical model for the ejected material of a Type IIb SN and, second, we describe our model of a shocked clump or knot in Cas A.  

\subsection{The type IIb SN ejecta }
\label{ejec}

\begin{table*}
\caption{Ejecta temperature T$_{gas}$ and number density n$_{gas}$ for the Type IIb Cas A supernova\tablefootmark{a} as a function of post-explosion time and ejecta zones (the mass coordinates of each zone in the He-core is given). }
\label{tab4}   
\centering
\begin{tabular}{c c c c c c c c c}
\hline
\hline

  & & Zone 1a & Zone 1b & Zone 2& Zone 3a & Zone 3b & Zone 4a & Zone 4b\\ 
 & & 1.68-1.88 \Ms& 1.89-2.18 \Ms& 2.19-3.86  \Ms& 3.87-4.02  \Ms& 4.03-4.49  \Ms& 4.5-5.26  \Ms& 5.27-5.62  \Ms\\
 Time & T$_{gas}$ & n$_{gas}$ & n$_{gas}$ &n$_{gas}$ & n$_{gas}$&n$_{gas}$ & n$_{gas}$& n$_{gas}$\\
\hline
100& 6664 & 7.2(7) & 1.4(8) & 1.8(8) & 2.0(8) & 3.0(8) & 7.2(8) & 7.7(8) \\
200& 2700 & 9.0(6) & 1.7(7) & 2.3(7) & 2.5(7) & 3.7(7) & 9.4(7) & 9.6(7) \\
300& 1600 & 2.7(6) & 5.0(6) & 6.6(6) & 7.5(6) & 1.1(7) & 2.8(7) & 2.8(7) \\
400& 1100 & 1.1(6) & 2.1(6) & 2.8(6) & 3.2(6) & 4.6(6) & 1.2(7) & 1.2(7) \\
500&   820 & 5.8(5) & 1.1(6) & 1.4(6) & 1.6(6) & 2.3(6) & 6.0(6) & 6.0(6) \\
1000& 330 & 7.2(4) & 1.4(5) & 1.8(5) & 2.0(5) & 3.0(5) & 7.5(5) & 7.8(5) \\
1500& 200 & 2.2(4) & 4.0(4) & 5.3(4) & 6.0(4) & 8.8(4) & 2.2(5) & 2.3(5) \\
2000& 140 & 9.0(3) & 1.7(4) & 2.3(4) &2.5(4) & 3.7(4) & 9.4(4) & 9.6(4) \\
2500& 100 & 4.6(3) & 8.7(3) & 1.1(4) & 1.3(4) & 1.9(4) & 4.8(4) & 4.9(4) \\ 
3000&   80 & 2.7(3) & 5.0(3) & 6.6(3) &7.5(3) & 1.1(4)&  2.8(4) & 2.8(4) \\ 
\hline 
\end{tabular}
\tablefoot{Time is in days, T$_{gas}$ in Kelvin and n$_{gas}$ in \cmc. \\
\tablefoottext{a}{The Type IIb SN ejecta is characterised by the following parameters: progenitor mass $=19$ \Ms; explosion energy~E$_{kin}= 1\times 10^{51}$~erg}; He-core mass~$=5.62$~\Ms; ejecta velocity $=4000$ \kms; \gray~Optical depth $\tau_{\gamma} = 15$.}
\end{table*}
Stars with mass in the range $8-30$ M$_{\odot}$ explode as core-collapse SNe of Type II, showing strong hydrogen lines in their spectrum. Type IIb SNe show a weak hydrogen line, that subsequently disappears with time, and are characterised by higher expansion velocities, resulting in a more diffuse ejected material compared to regular Type II SNe. The Cas A remnant ensues from the explosion of a blue supergiant of mass $15-25$ \Ms\ as a type IIb SN (\cite{krau08}). 

We consider a stratified ejecta whose elemental composition is given by the 19 \Ms\ SN progenitor model of Rauscher et al. (2002). The ejecta consists of mass zones of specific chemical composition summarised in Table \ref{tab3}. Each zone is microscopically mixed and we assume no chemical leakage between zones. 

The number density and temperature profiles are taken from explosion models for Type IIb SNe presented by Nozawa et al. (2010), where a homologous expansion is assumed for the ejecta gas, at constant velocity $v_{ej}=4000$ \kms\ (i.e., the expansion velocity of the oxygen-rich core in Nozawa's model). The gas number density is given by 
\begin{equation}
\label{eq3}
n (M_r, t) = n(M_r, t_0) \times {\left(\frac{t}{t_0}\right)}^{-3},
\end{equation}
where $M_r$ is the mass zone position, and $t$ the time after explosion with $t_0=100$ days. The temperature profile is given by 
\begin{equation}
\label{eq4}
T(t) = T_0 \times {\left(\frac{t}{t_0}\right)}^{3(1-\gamma)}
\end{equation}
where $T_0$ is the gas temperature at 100 days, and $\gamma$ is the adiabatic index. Values are: $T_0=6663.95$ K and $\gamma = 1.433$. A flat density profile was assumed across the He-core with a value at 100 days equal to $\rho(100)= 5.1\times 10^{-15}$ g~\cmc. All values were taken from Nozawa et al. (2010). The SN Type IIb parameters for the Cas A progenitor and the values for the ejecta gas number density and temperature as a function of time are summarised in Table \ref{tab4}.  


\subsection{The reverse shock models}
\label{revsh}

\begin{figure}
   \centering
   \includegraphics[width=\columnwidth]{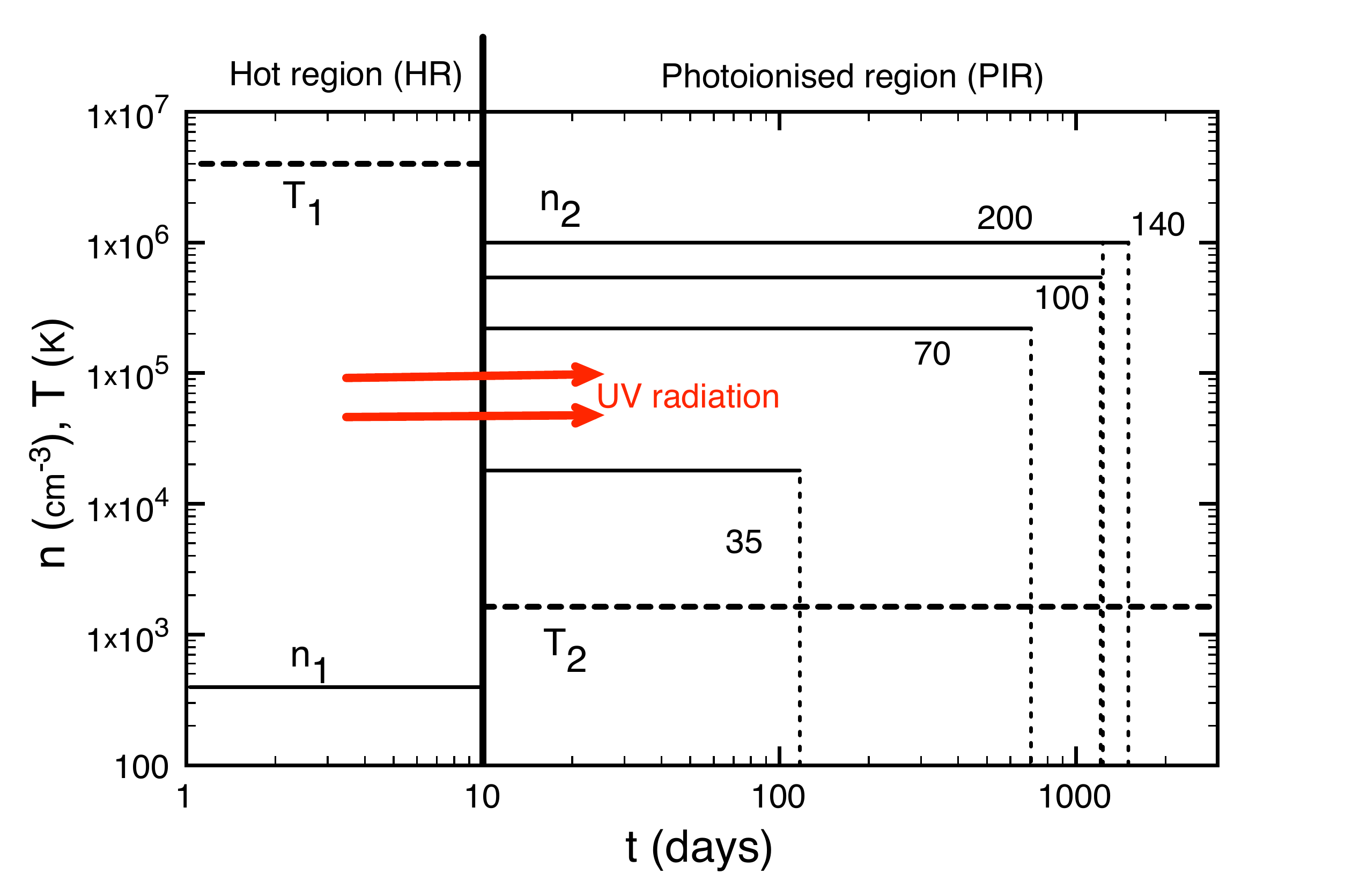}
      \caption{Schematic post-reverse shock structure, including the cooling hot region (HR) and photoionised region (PIR). The pre-shock values are $T_0$ and $n_0$. Gas temperature (dashed line) and number density (solid line) are given in Table \ref{tab5}. Number density and PIR duration are indicated for the RS velocities studied in this paper. }
\label{fig1}
\end{figure}

The Cas A remnant shows evidence for several high-density knots or clumps embedded in a rarefied inter-clump medium. Fesen et al. (2001) have observed fast-moving knots (FMKs) in the optical while Rho et al. (2009, 2012) have detected ro-vibrational transitions of CO in emission in several small knots located at the RS position. High energy line detection of shocked CO with Herschel confirmed the existence of high-density gas regions in the remnant (\cite{wal13}). In Cas A, the RS velocity relative to the ejecta is \env\ 2000 \kms\ (\cite{mor04}). When encountering a dense clump with a density contrast with respect to the inter-clump medium $\chi= n_c/n_{ic}$ ($n_c$ and $n_{ic}$ are the number density of the clump and the interclump medium, respectively), the RS velocity decreases by a factor $\sqrt\chi$ owing to the conservation of energy. For clump density contrast ranging from between $100$ and $1000$ and RS velocities of $1000-2000$ \kms, the RS velocity in the clump spans $30-200$ \kms. Such velocities have been confirmed by DS10 from their analysis of fine-structure far-IR atomic lines coming from an oxygen-rich FMK in Cas A. 

In the present study, we want to assess the fate of the chemical species, including molecules and dust clusters, within ejecta clumps that are processed by the RS. We thus investigate various RS velocities, derive the time-variation of the post-RS gas parameters, based on existing shock models, and apply our chemical kinetic formalism to study the post-RS chemistry. Borkowski \& Shull (1990) (hereafter BS90) model steady state radiative shocks in a pure oxygen gas with velocities that range between 35 \kms\ and 170 \kms. This study provides the adequate range of conditions for the case of the RS impacting a dense knot in Cas A. 

Considering a knot of gas with an initial density contrast with respect to the inter-clump medium, the parameters pertaining to a description of the RS, as illustrated in Figure \ref{fig1}, are the shock velocity relative to the knot velocity $V_s$, the pre-shock number density and temperature of the knot, $n_0, T_0$, the gas density and temperature in the post-shock hot region (HR), $n_1, T_1$, and those for the photoionised region (PIR), $n_2, T_2$, and the time length of the PIR, $t_{PIR}$. In the HR, the compression induces an equilibration between ions and electrons which draw their temperature through Coulomb collisions with ions when they cool via inelastic collisions with ions. The cooling time being shorter than the recombination time for ions, a constant ionisation state prevails in the HR until the gas quickly cools down and recombines. Ionisation is still sustained in the PIR by the UV flux coming form the HR so that thermal equilibrium is reached over a time $t_{PIR}$ (or length $L_{PIR}$) defined by the depletion of the HR UV flux. 

From our SN ejecta model, we can constrain the knot physical properties as follows: considering the oxygen-rich zones of the ejecta (e.g., zone 2) and assuming the ejecta expansion velocity of Table \ref{tab4}, the volume of the homogenous ejected material  at time $t$ (in days) after explosion, assuming a spherical expansion, is
\begin{equation}
\label{eq5}
 V(t) = {{4\times \pi}\over {3}} \times (v_{ej}\times t \times 8.64\times 10^4)^3, 
\end{equation}
where $v_{ej}$ is the ejecta velocity. For our chosen values at day 100, $V(100) = 1.7 \times 10^{47}$ \cmc. Values for the clump volume filling factor, $f_c$, derived from radiative transfer studies of the dust emission in SN ejecta typically range from 0.05 to 0.2 (\cite{gal12}). Assuming $f_c = 0.05$, we derive that the total volume in the form of clumps in the SN ejecta is $V_{tot, c}(100)= f_c \times V(100) = 8.6 \times 10^{45}$ \cmc. For the ejecta mass given in Table \ref{tab4}, we derive a typical clump mass ranging from $10^{-3}$ \Ms\ to $3 \times 10^{-3}$ \Ms, where the number of clumps $N_c$ is 1000 and 3000, respectively. These $N_c$ values are usually used in radiative transfer models. Our derived clump mass is also consistent with values obtained from 3D hydrodynamic models of SN explosion (\cite{ham10}). The number density of one oxygen-rich clump at day 100 post explosion is then given by

\begin{equation}
\label{eq6}
 n_c(100) = {M_c \over{V_{tot, c}(100)/ N_c}},
\end{equation}
with $n_c(100) = 3.4 \times 10^{10}$ \cmc\ for $N_c=1000$. This gas number density is roughly a factor of 200 larger than the values listed in Table \ref{tab4} for the O-rich zones. Using Eq. \ref{eq3}, we recover a typical oxygen-rich clump number density 330 years after explosion of \env\ $20$ \cmc, which is consistent with the value of $100$ \cmc\ inferred by DS10 for the clump density not yet shocked by the RS. If we assume an inter-clump medium number density of $1$ \cmc\ and $0.1$ \cmc (\cite{noz03}, BS90), we recover a clump/inter clump density contrast $\chi$ value of between \env\ 20 and 200, respectively. The density contrast used by DS10 is $100$ and that assumed by Sylvia et al. (2012) is $1000$. For a range of initial unattenuated RS shock speeds of $1000 - 2000$ \kms, the RS velocity through the clump ranges from \env $35$ \kms\ to $200$ \kms. These RS velocities have been modelled by BS90 and we use their results as a basis for our RS analytical models.

\begin{table}
\caption{Reverse shock parameters as a function of attenuated RS velocity. The pre-shock density $n_0$ and the PIR length $L_{PIR}$ for all PIR models are 100 \cmc\ and $2\times$10$^{11}$ cm, respectively. }           
\label{tab5}       
\centering                           
\begin{tabular}{l c c c c c}         
\hline\hline                  
$V_s$ & Log$N(O)$\tablefootmark{a} & $n_2$& $T_2$\tablefootmark{b} &$v_2$ & $t_{PIR}$ \\
\hline                        
35 & 15.6 &$1.8\times10^4$ & 182 & 194.5 &0.33/120\\
70 & 16.6 & $2.2\times10^5$ & 1500 & 31.8& 1.98/723 \\
100 & 17.0& $5.4\times 10^5$ & 1500 & 18.5 &3.4/1241 \\
140 & 17.3& $1\times10^6$ & 1500 & 14.0& 4.21/1537 \\
200\tablefootmark{c} & 17.0 & $1\times10^6$ & 1500 & 20.0& 3.17/1157 \\
\hline                                    
\end{tabular}
\tablefoot{Units - $V_s$: \kms; $N(O)$: cm$^{-2}$; $n_2$: \cmc; $T_2$: K; $v_2$: m s$^{-1}$; $t_{PIR}$: year/days.\\
\tablefoottext{a}{Values for the  column density $N(O)$ are taken from the shock models with conduction (70, 100, 140 \kms) and without conduction (35 \kms) of BS90 except for the 200 \kms\ case.}
\tablefoottext{b}{The T value for the 35 \kms\ shock is from BS90 while other values are derived from DS10.}
\tablefoottext{c}{Values for this model are derived from DS10 except for the UV flux taken from BS90 for the RS velocity 170 \kms.}}
\end{table}
 
The RS parameters are listed in Table \ref{tab5}. We assume five velocities for the shock relative to the clump, an initial pre-shock density in the clump of $n_0=100$ \cmc\ and a PIR length $L_{PIR} = 2\times 10^{11}$ cm for all models, as derived by DS10. In the PIR, we assume the oxygen column densities listed in Table 10 of BS90 for the various shock speeds listed in Table \ref{tab5}. The time length of the PIR is derived assuming the conservation of momentum through the shock front and is given by 

\begin{equation}
\label{eq7}
 t_{PIR} = {N(O) \over{n_2\times v_2}} = {N(O) \over{n_0\times V_s}},
\end{equation}
where $N(O)$ is the oxygen column density, and $v_2$ the gas velocity in the PIR. The density in the PIR $n_2$ is derived from the definition of the PIR length $L_{PIR}$ given by
\begin{equation}
\label{eq8}
 L_{PIR} = v_2 \times t_{PIR} = {n_0 \over n_2} \times V_s \times t_{PIR}. 
\end{equation}
The temperature in the PIR is that derived by DS10 for the high velocity shocks and is fixed to $1500$~K. We also test the impact on the PIR chemistry of the higher PIR temperature derived by BS90 ($T_2=4500$ K). For the slowest shock, we take the PIR temperature derived for that shock speed by BS90. 

Non-thermal processes include the penetration of the PIR by UV photons coming from the HR. These UV photons partially sustain the level of ionisation in the PIR via the photoionisation of oxygen. BS90 estimate the photon fluxes emergent from the HR as a function of the shock velocity and energy band. We consider the ionisation of OI in various levels of OII involving UV photons with energy from between 13.62 and 35.12 eV as the main source of electrons in the PIR. UV flux values are listed in Table 5 of BS90 for the shock speeds considered in this study. The unattenuated photoionisation rate for oxygen $\zeta_{0,\nu}$ at frequency $\nu$ is given by 

\begin{equation}
\label{eq9}
\zeta_{0,\nu} = \int_{\nu} \sigma_{\nu} \times 4 \pi \frac{J_{\nu}}{h\nu}\times d\nu, 
\end{equation}

where 4$\pi$J$_{\nu}$ is the radiation intensity averaged over the solid angle and $\sigma_{\nu}$ is the photoionisation cross section of oxygen taken from \cite{an88} at the frequency $\nu$. The attenuated rate due to extinction in the post-shock gas is calculated as follows: we omit potential attenuation from dust grains present in the PIR (this will be considered in paper II.). We estimate the unattenuated rate for the various frequency bands $k$ and the integrated fluxes given in Tables 3 and 5 of BS90, respectively, and calculate the frequency-dependent optical depth $\tau_k$ for each frequency band $k$, which is given by 

\begin{equation}
\label{eq10}
\tau_k(t) =  \sigma_k \times v_2\times t \times n_2. 
\end{equation} 
where $t$ is the time in the PIR. The total attenuated photoionisation rate at time $t$ in the PIR is then given by 

\begin{equation}
\label{eq11}
\zeta(t) = \sum_k \zeta_{0,k}\times  exp [ - \tau_k(t)],
\end{equation}

where $\zeta_{0, k}$ is the unattenuated photoionisation rate in the frequency band $k$. The unattenuated photoionisation of oxygen is then included as a non-thermal process in the chemical scheme described in \S \ref{chem} with the rate calculated from Equation \ref{eq9} for the frequency bands $k$, and the attenuated rate is calculated at each integration time step according to Equation \ref{eq11}.

      \begin{figure}
   \centering
   \includegraphics[width=\columnwidth]{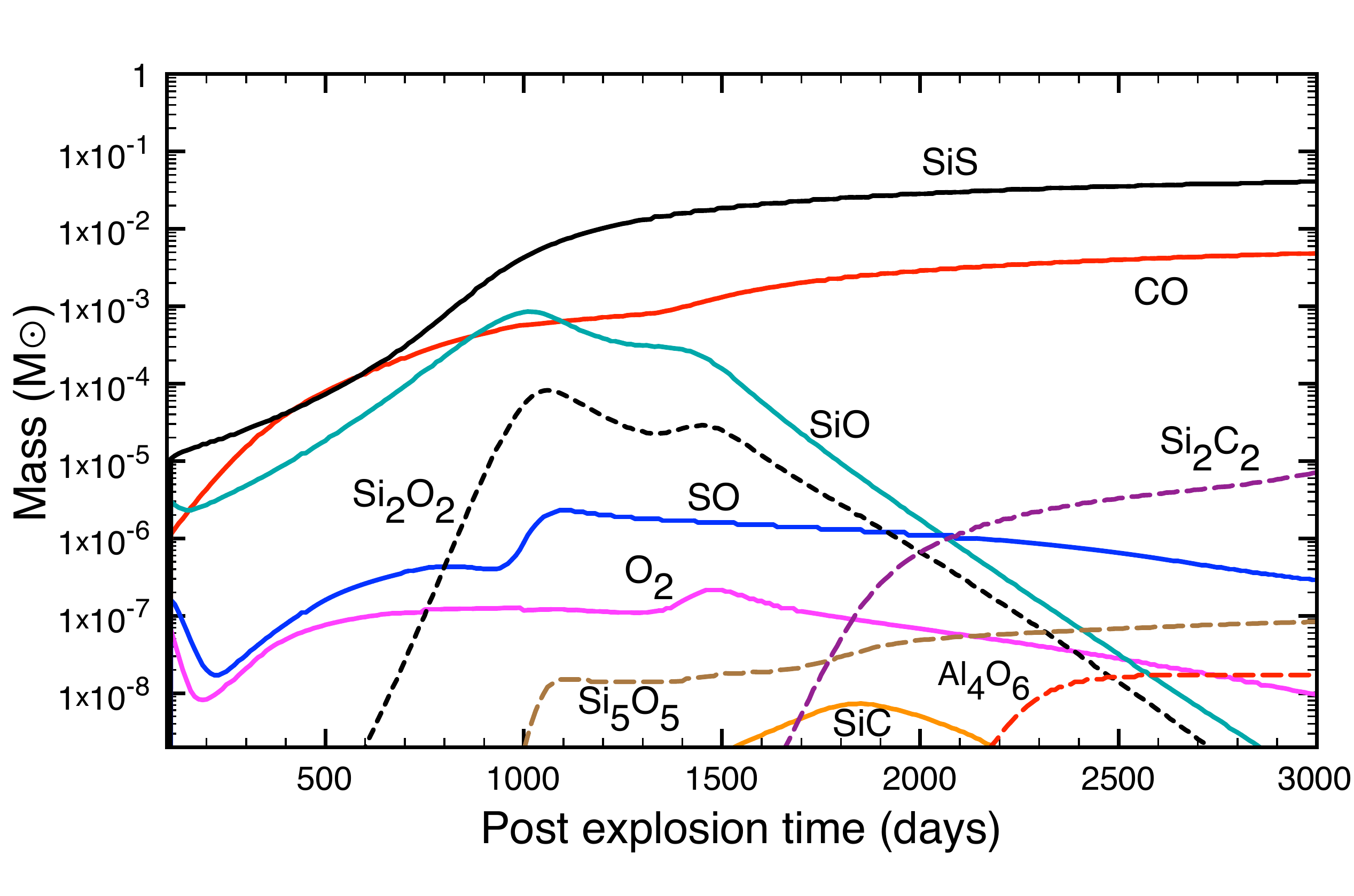}
 \caption{Masses of molecules and dust clusters in the ejecta as a function of post-explosion time for the Type IIb SN with 19~\Ms\ progenitor that led to Cas A. The ejecta parameters are given in Table \ref{tab4}. Because of the low ejecta number density, no significant amount of dust clusters is formed.}
 \label{fig2}
   \end{figure}

\section{Results}
\label{res}

We study the variation of the abundances and masses of chemical species and dust clusters that form in the ejecta of the Cas A supernova precursor and present the results in \S \ref{rescasa}. We then focus on an oxygen clump in Cas A whose chemical composition has been derived in \S \ref{rescasa}, and study the impact of the reverse shock on such a dense clump. Various reverse shock velocities that correspond to clump/interclump density enhancement of $100-1000$ are considered and results are presented in \S \ref{resRS}.

\subsection{Molecules and dust clusters in the Cas A supernova progenitor }
\label {rescasa}

Type IIb SNe have low-density ejecta compared to their type II-P counterparts. In \S \ref{lowden}, we present the results for a stratified ejecta whose parameters are given in Table \ref{tab4}. We define this low-density Type IIb SN ejecta as our "standard case". Results on the impact of increasing the gas density on the chemical composition of the gas and the dust are presented in \S \ref{highden}. 

\subsubsection{Low density Type IIb ejecta} 
\label{lowden}

The masses of the prevalent molecules and dust clusters that form in all the ejecta zones as a function of post-explosion time are shown in Figure \ref{fig2} for the "standard case". The dominant species is SiS which efficiently forms in zone 1A/B of the ejecta from the radiative association reaction 

\begin{equation}
\label{R1}
Si + S \longrightarrow SiS + h\nu.
\end{equation} 

This reaction has already been identified as the prevalent formation process for SiS in the innermost zones of the denser ejecta of Type II-P SNe (SC13). This reaction is temperature-independent and can therefore proceed at late post-outburst time and ensure the growth of SiS mass with time. SiS destruction is provided by the reaction with \arp. CO forms in the oxygen-rich zones 3A/B and 2 via the radiative association reaction 
\begin{equation}
\label{R2}
C + O \longrightarrow CO + h\nu, 
\end{equation} 
and the reaction with O$_2$,
\begin{equation}
\label{R3}
C + O_2 \longrightarrow CO + O. 
\end{equation} 
CO destruction stems from charge-exchange reaction with O$^+$ and with reaction with Ne$^+$. Silicon monoxide, SiO, primarily forms in zones 1B and 2, with a small contribution from zones 3A/B,  from the radiative association process 
\begin{equation}
\label{R4}
Si + O \longrightarrow SiO + h\nu.
\end{equation} 
and is destroyed by the reaction with \nep. Its abundance increases until \env\ day 1300 when SiO starts to be depleted in SiO dimers, \sidim, and trimers. Finally, SO and O$_2$ also form from radiative association reactions in zone 1A and 1B/2/3, respectively. Inert gas ions do not efficiently recombine at early time because of the low ejecta gas density and their abundance gradually decreases owing to the decrease in \Ni~mass and \gray~photons with time. The formation of the prevalent molecules is then postponed to $t> 1000$ days, in contrast with Type II-P SNe (SC13). 

The masses of the most abundant synthesised dust clusters are also shown in Figure \ref{fig2}. A small amount of Si$_5$O$_5$ is formed in the O-rich zones 1B, 2, and 3A/B out of SiO and its polymerisation. The dimerisation of SiO is possible after day 1000 but further growth is inhibited owing to the very low ejecta gas density. No significant amount of silicate dust clusters is formed in the O-rich zones of the ejecta because the first chemical steps in the synthesis of forsterite and enstatite dimers involve neutral-neutral processes with a moderate activation barrier (\cite{gou12}, SC13), and higher gas temperatures than those found at day \env~1000 in the ejecta are thus required. The low gas density at day $> 1000$ also hampers the efficient nucleation of silicate clusters. Some SiC dimers form in the carbon-rich outermost zones 4A/B after day 1800, but as for SiO polymers, their later growth is hampered by the low gas density at late times. The masses of molecules and dust clusters at 3000 days after explosion and for the standard case are summarised in Table \ref{tab6}. We see that the mass fraction of the ejected material in the form of molecules only amounts to $0.2$ \% and that dust clusters form in small amount. Therefore, the low-density, homogeneous ejecta of a Type IIb SN seems to be inefficient at forming significant quantities of dust clusters and grains. 

\begin{table}
\caption{Masses of molecules and dust clusters (in \Ms) at 3000 days post-explosion for the Type IIb SN with a 19~\Ms\ progenitor that led to Cas A. The gas density cases correspond to the "standard case" and increase in gas number density of 200 and 2000, as discussed in the text. Efficiencies are the mass fraction of the ejected material in the form of molecules or dust clusters. }           
\label{tab6}       
\centering                           
\begin{tabular}{l ccc}         
\hline\hline                  
 Density Case & Standard & x200 & x2000 \\
 \hline
Molecules &  & & \\ 
\hline                        
SiS& $4.1\times 10^{-2}$  &$1.3\times10^{-1}$& $5.7\times 10^{-2}$\\
CO & $4.8\times 10^{-3}$ & $4.9\times10^{-1}$ & $4.9\times 10^{-1}$\\
SiO& $6.7\times 10^{-10}$& $8.7\times 10^{-12}$& $1.8\times 10^{-12}$ \\
O$_2$ & $9.5\times 10^{-9}$& $5.8\times10^{-1}$ & 1.07 \\
SO & $2.9\times 10^{-7}$ & $2.6\times10^{-2}$& $7.7\times 10^{-2}$ \\
Total Mass&$4.6\times 10^{-2}$ & 1.22& 1.69\\
Efficiency &1.2\% & 31 \% & 47 \%\\
 \hline
Clusters&  &  &\\ 
\hline                        
Silicates\tablefootmark{a}&$ -- $ &$3.2\times10^{-3}$ &$1.3\times 10^{-2}$\\
SiC & $7.9\times 10^{-6}$ & $3.5\times10^{-4}$& $1.3\times 10^{-4}$  \\
Silica\tablefootmark{a}& $2.7\times 10^{-7}$& $1.4\times 10^{-9}$& $8.0\times 10^{-15}$  \\
Alumina &$1.7\times 10^{-8}$ &$1.7\times 10^{-2}$& $1.4\times 10^{-2}$ \\
Pure metals\tablefootmark{a}& $ 4.5\times 10^{-10}$ & $8.8\times10^{-5}$& $8.0\times 10^{-3}$ \\
Carbon & $ -- $ & $7.5\times10^{-12}$& $9.4\times 10^{-3}$\\
Iron sulphide & $ -- $& $6.6\times 10^{-10} $& $6.2\times 10^{-7}$ \\
Total Mass &$8.2\times 10^{-6}$ &$0.021$&0.045 \\
Efficiency &0.0003 \% & 0.52 \% & 1.1\%\\
\hline
\end{tabular}
\tablefoot{
\tablefoottext{a}{Silicates correspond to forsterite clusters, pure metals to Fe, Si, and Mg clusters, and silica to polymers of SiO.}}
\end{table}
 
\begin{figure}
 \centering
\includegraphics[width=\columnwidth]{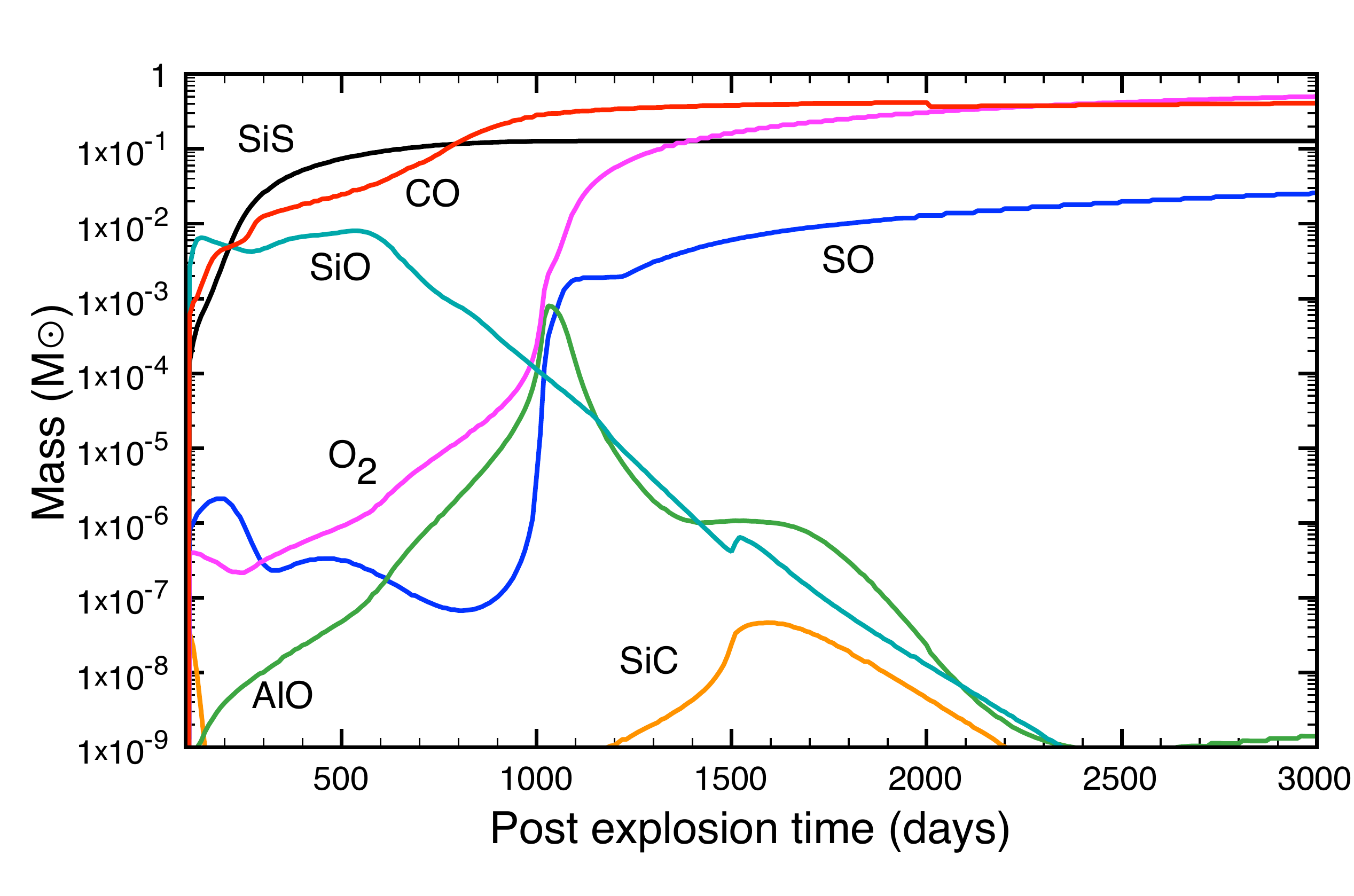}\\
\includegraphics[width=\columnwidth]{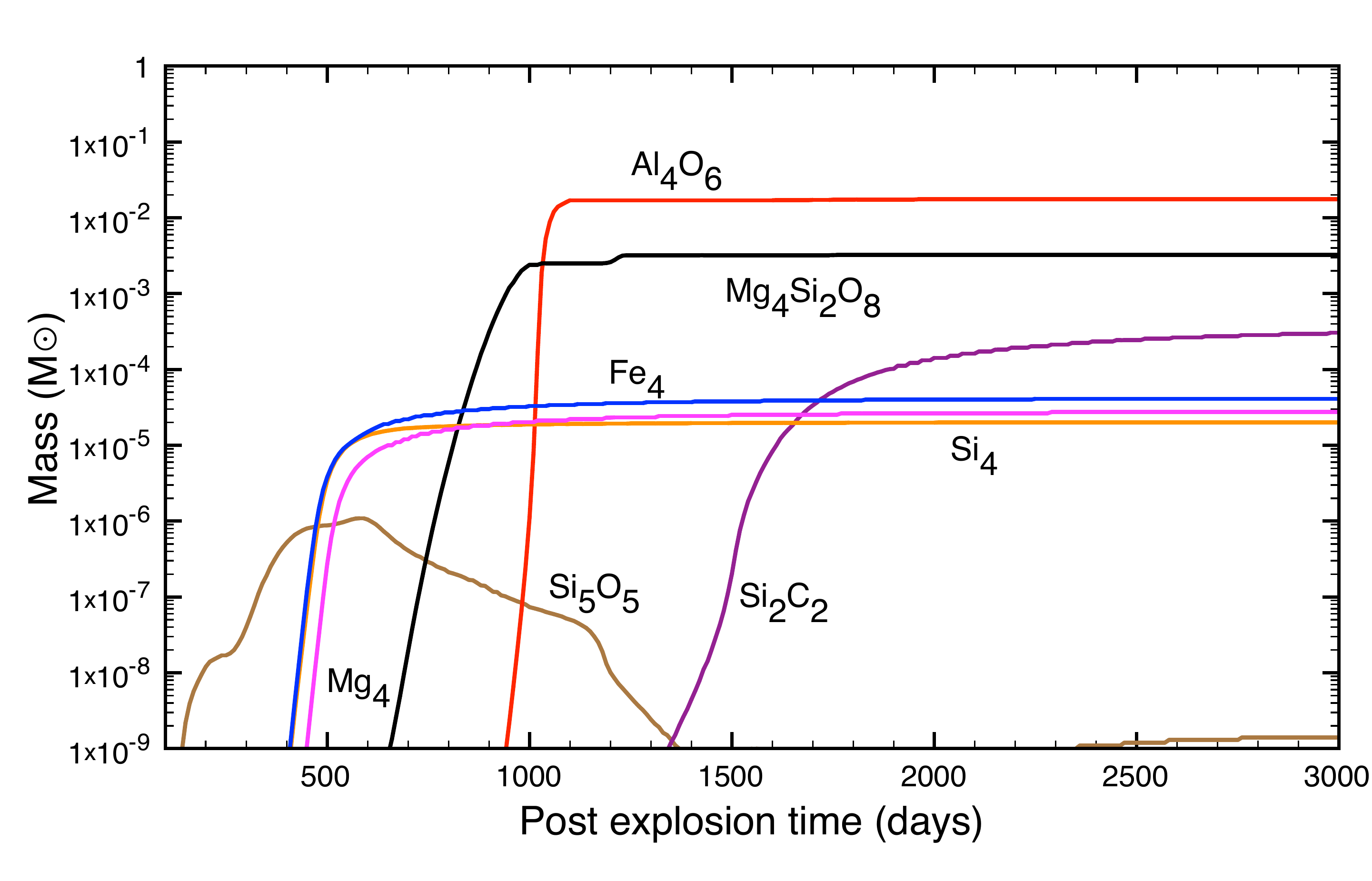}
  \caption{Masses of molecules and dust clusters in the ejecta as a function of post-explosion time for the Type IIb SN with 19~\Ms\ progenitor that led to Cas A, assuming a number density increase of 200 with respect to the "standard case". Top: masses of molecules; Bottom: masses of dust clusters. }
\label{fig3}
     \end{figure}
These results are in contrast with the theoretical study of dust formation in the SN progenitor of Cas A presented by Nozawa et al. (2010), where the authors derive a total dust mass of $0.167$~\Ms\ for similar ejecta conditions and a complex chemical composition for the dust, including silicates, carbon and metal oxides. Their study of dust synthesis is based on a CNT formalism and assumes that all available carbon and silicon atoms are locked up in CO and SiO when the C and Si elemental abundances are less than that of oxygen. The formalism fails to include the chemistry of the ejecta gas phase, and the formation and destruction of molecules and dust clusters from chemical kinetics. We find that only \env~$1\times 10^{-5}$~\Ms\ of silicon carbide, silica, and alumina dust clusters can form for the standard case. Our results point to the importance of the nucleation phase, which describes the formation of molecules and dust clusters out of the elements comprised in the ejecta zones, as a bottleneck to dust condensation. This issue has already been addressed by Cherchneff \& Lilly (2008),  Cherchneff \& Dwek (2009, 2010) and SC13 and will be discussed in \S \ref{dis}. On the observational front, Spitzer data of Cas A reveal that the IR excess due to dust in the remnant can be reproduced by an ensemble of dust grains of various composition, including silicates, metal oxides and sulphides, pure metal grains and some amorphous carbon grains (\cite{rho08, ar14}). Dust is clearly present in the diffuse remnant phase, implying that it could form in the denser ejecta of the supernova. This is in contrast with our present findings and forces us to reconsider the impact of the gas density on the nucleation of dust. 

\subsubsection{Impact of gas density on the chemical composition of the ejecta}
\label{highden}

We now increase the ejecta number density of our standard case by various factors $x$ ranging from between 10 and $2000$ (the standard case corresponds to $x=1$), and study the nucleation phase and the chemical composition of the gas. This increase is applied to all zones of the ejecta at day 100 and the gas number density follows the time dependence given by Equation \ref{eq3}. We do not aim at modelling a clumpy SN ejecta but want to study the impact of density enhancements that characterise clumps on the ejecta chemistry. 
According to \S \ref{revsh}, clumps are characterised by a number density roughly 200 times larger than that of the homogeneous ejecta given in Table \ref{tab4}. Results on the species masses for a density enhancement of $200$ are then presented in Figure \ref{fig3}. The masses of both molecules and dust clusters have significantly increased. The chemical processes responsible for the formation of the prevalent molecules vary from those reported in the previous section because of the high gas density. While radiative association reactions contribute to the synthesis of all molecules in the ejecta, bimolecular processes including reactions with O$_2$ contributes in forming CO, SO, and AlO in the various zones. SiS is formed essentially in the innermost zones 1A/B. As shown by SC13 for Type II-P SNe, CO is mostly synthesised in the O-rich zones 2 and 3A/3B and does not trace the formation of carbon dust clusters. SiO forms essentially in zone 1B and 2 and gets quickly depleted into silicate clusters after day \env~600. The dust clusters start to form as early as 400 days post-explosion for the pure metal clusters, and significant masses of alumina and forsterite clusters form in the O-rich zones after day 800. 
         \begin{figure}
   \centering
   \includegraphics[width=\columnwidth]{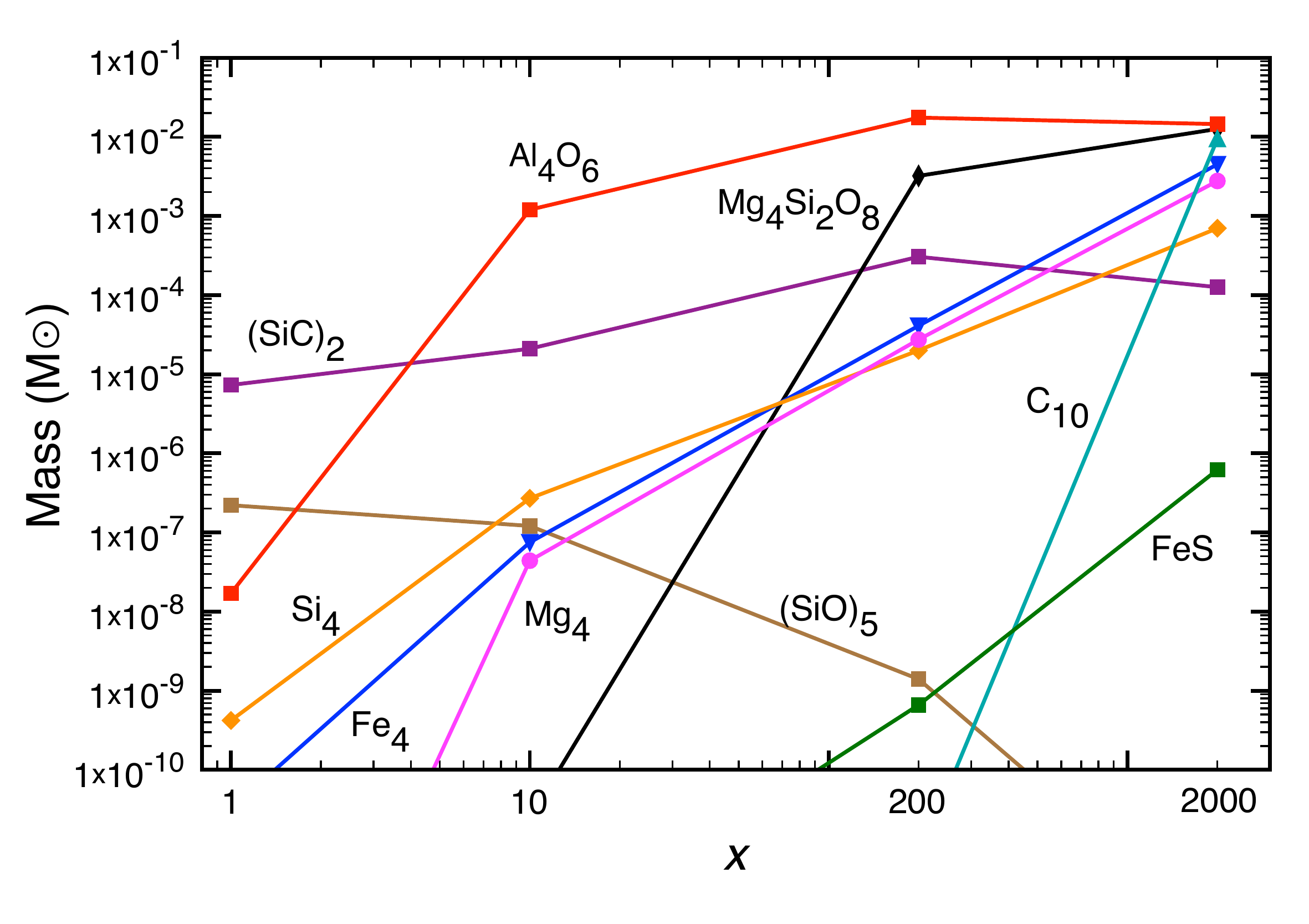}
      \caption{Masses of dust clusters in the ejecta at 3000 days post-explosion as a function of density increase for the Type IIb SN with 19 \Ms\ progenitor that led to Cas A. The $x$ axis represents the increment of density with respect to our "standard case" (which corresponds  to $x=1$). The chemical complexity of dust clusters grows with increasing ejecta number density to reach a composition comparable to that of the dust formed in Type II-P SNe for $x=2000$ (\cite{sar13}).   }
 \label{fig4}
   \end{figure}
   

The masses of dust clusters at day 3000 are shown in Figure \ref{fig4} as a function of gas number density increase in the ejecta. The build-up of the chemical complexity of the dust in the ejecta is clearly seen. As seen in the previous section, small amounts of SiC, silica, and alumina dust clusters form in the "standard case" ejecta given by the explosion model of a Type IIb SN, with upper limit on the dust mass not exceeding \env\ $ 1\times 10^{-5}$~\Ms. More complex dust types are synthesised as the density increases, with the formation of alumina,  SiC, and pure metal clusters (silicon, iron, and magnesium) when the density is raised by a factor of 10. For a density enhancement factor of 200, the prevalent dust clusters are alumina and forsterite, with SiC and pure metal clusters. The composition of the dust formed for the enhancement factor of 2000 resembles that derived for a SN of type II-P and prevalently includes alumina and forsterite (SC13). The carbon dust clusters seem to be the most density-dependent clusters and start forming for the largest increase in the density. In a high density media, carbon dust clusters form in the outermost zone of the ejecta from the initial production of the C$_2$ carbon chain, which grows via C and C$_2$ addition to form long carbon chains and the first cyclic ring C$_{10}$. These molecules form at late post-explosion time (\env 1000 days) once the \hep\ ion abundance drops owing to ion recombination and decrease in the \gray\ flux (SC13). In the case of carbon chains, the synthesis of C$_{10}$ involves many chemical steps and necessitates the high gas density to facilitate the recombination of \hep. For the enhancement factor $x= 200$, the gas density is not sufficient to weaken the \hep\ recombination and permits the survival of \hep\ ions at very late times. Therefore, almost negligible masses of carbon dust precursors are produced as seen in Figure \ref{fig4}. 

         \begin{figure}
   \centering
   \includegraphics[width=\columnwidth]{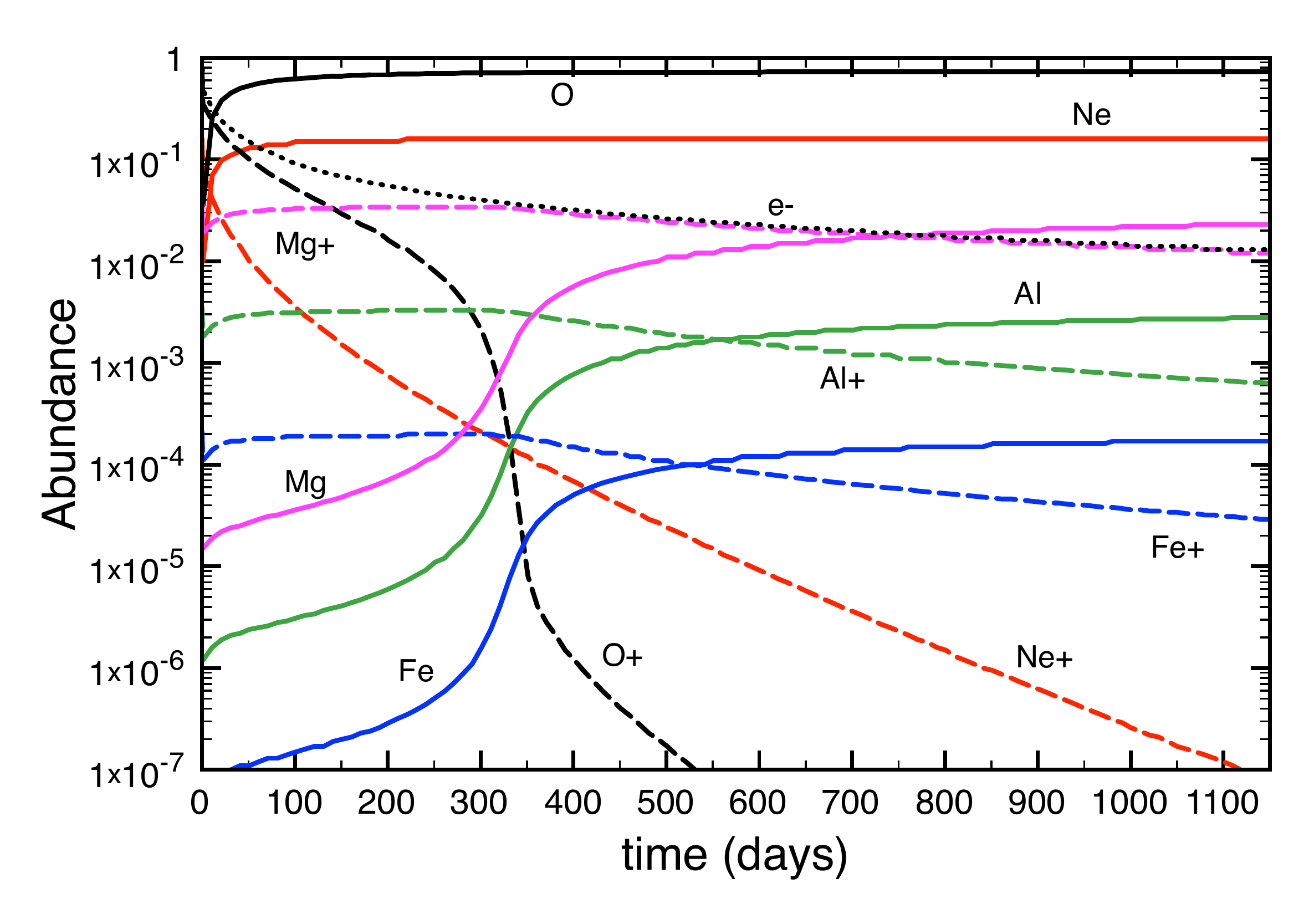}
    \includegraphics[width=\columnwidth]{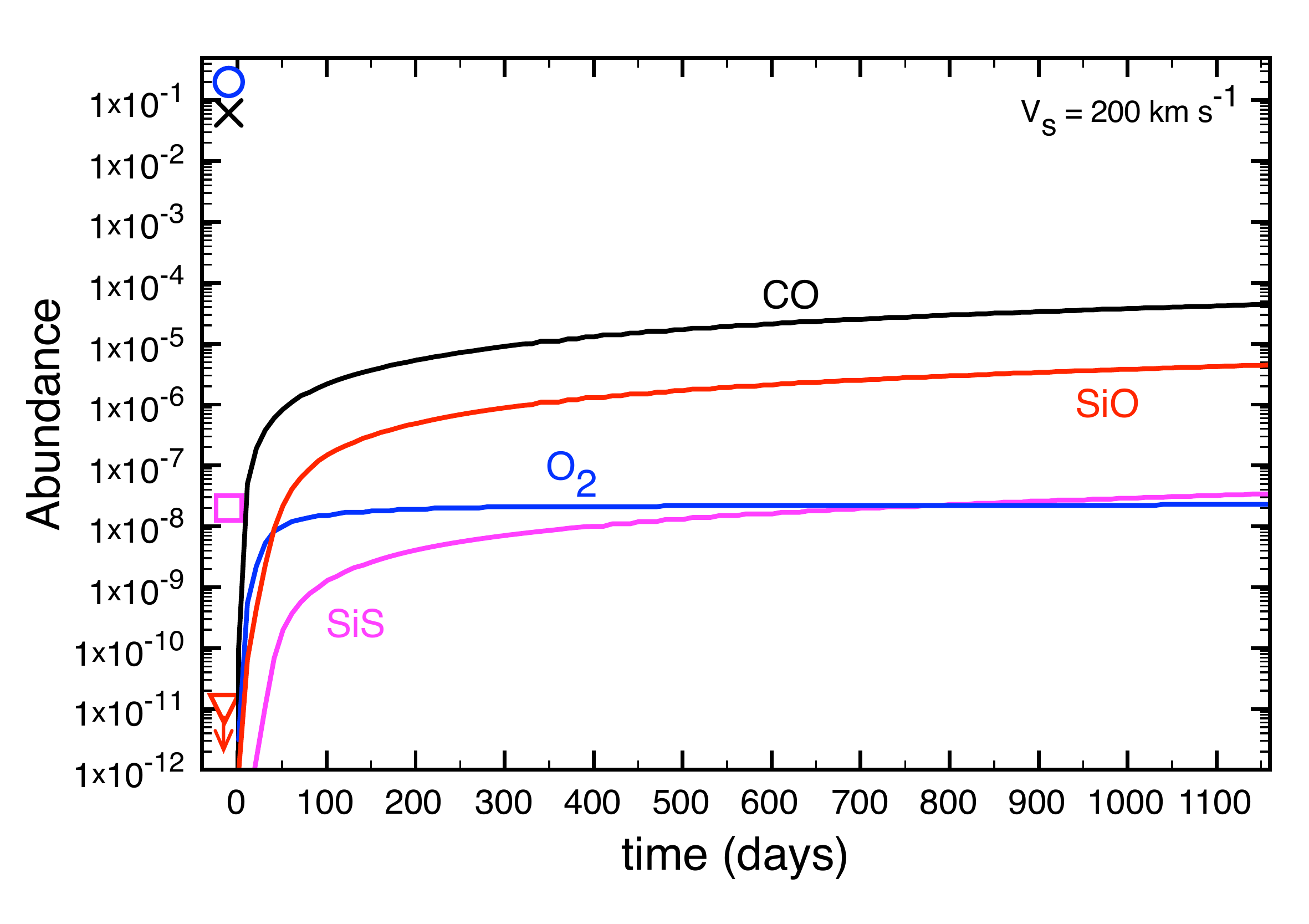}
      \caption{Abundances in the PIR (with respect to total gas number density) for the 200 \kms\ shock model. Top: Ions and atoms. Bottom: Molecules. Abundances in the pre-shock gas are those derived from \S 4.1.2 at 3000 days post-explosion and for a density increase $x=200$. They are plotted as symbols at $t=0$ (blue circle: O$_2$; black cross CO; pink square: SiS; red triangle: SiO). The SiO pre-shock abundance value is $1\times 10^{-17}$.   }
 \label{fig5}
   \end{figure}
      

The masses of molecules and dust for the cases $x=200$ and $x=2000$ are summarised in Table \ref{tab6} at day 3000 after explosion. We clearly see that the efficiency at forming molecules and dust clusters increases with gas density enhancement but species are affected in different ways. For example, O$_2$ and SO are more responsive to a rise in gas density owing to the larger number of chemical reactions involved in their formation and the fact that SO forms from O$_2$ at high density (SC13). More generally, we conclude that a low-density stratified ejecta as derived from non-clumpy explosion models of Type IIb SNe are not conducive to the synthesis of large amounts of molecules and are almost dust-free. High-density clumps thus need to be considered as the main sources of molecules and dust in the SN ejecta, to explain the presence of dust observed at IR wavelengths in Cas A. These clumps may retain the chemical composition of the ejecta zones from where they originate, evolve, and are shocked by the reverse shock in the remnant phase. 

\subsection{Molecules and dust clusters in the reverse shock}
\label{resRS}

\begin{figure*}
 \centering
\includegraphics[width=\columnwidth]{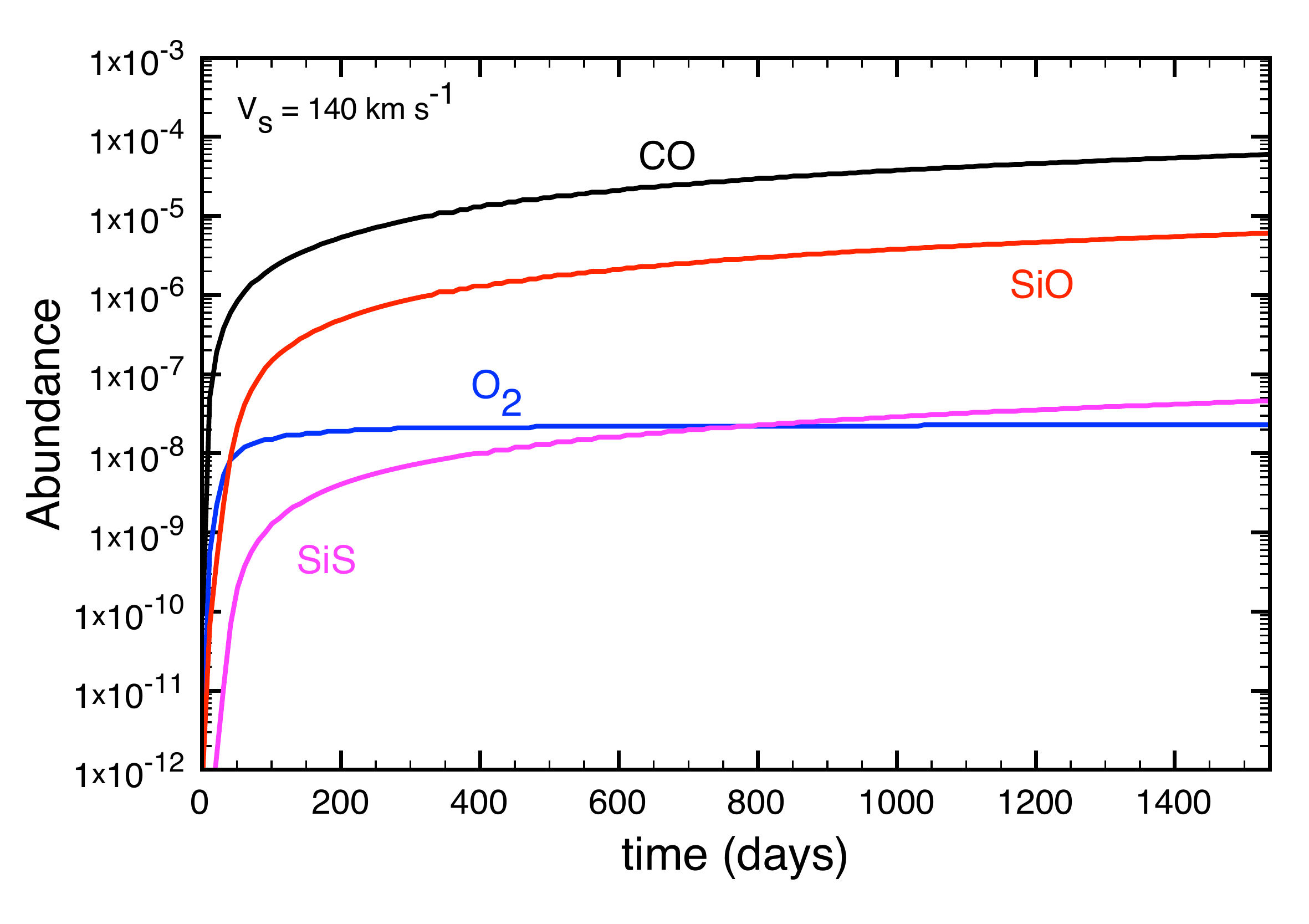}
\includegraphics[width=\columnwidth]{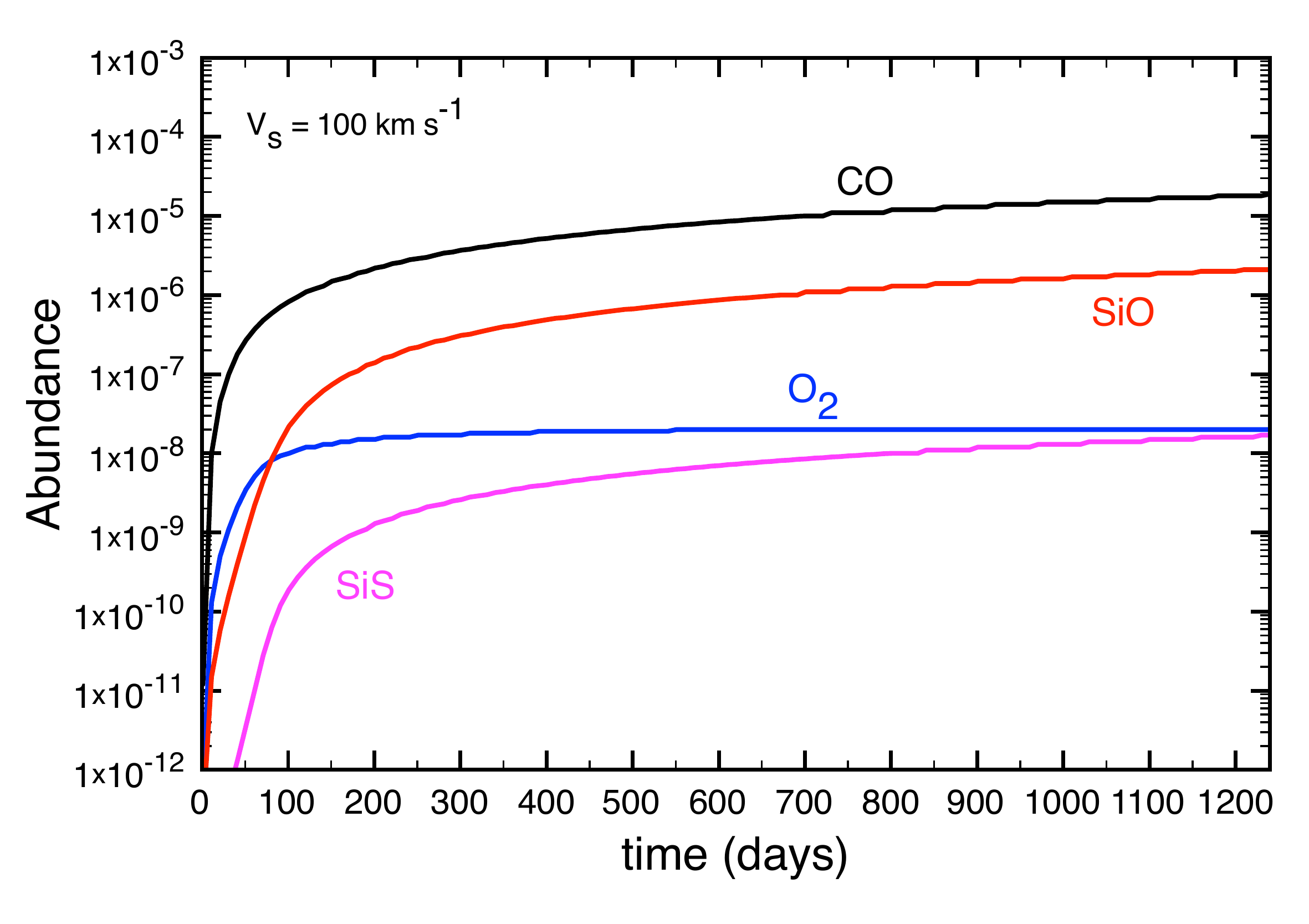}
\includegraphics[width=\columnwidth]{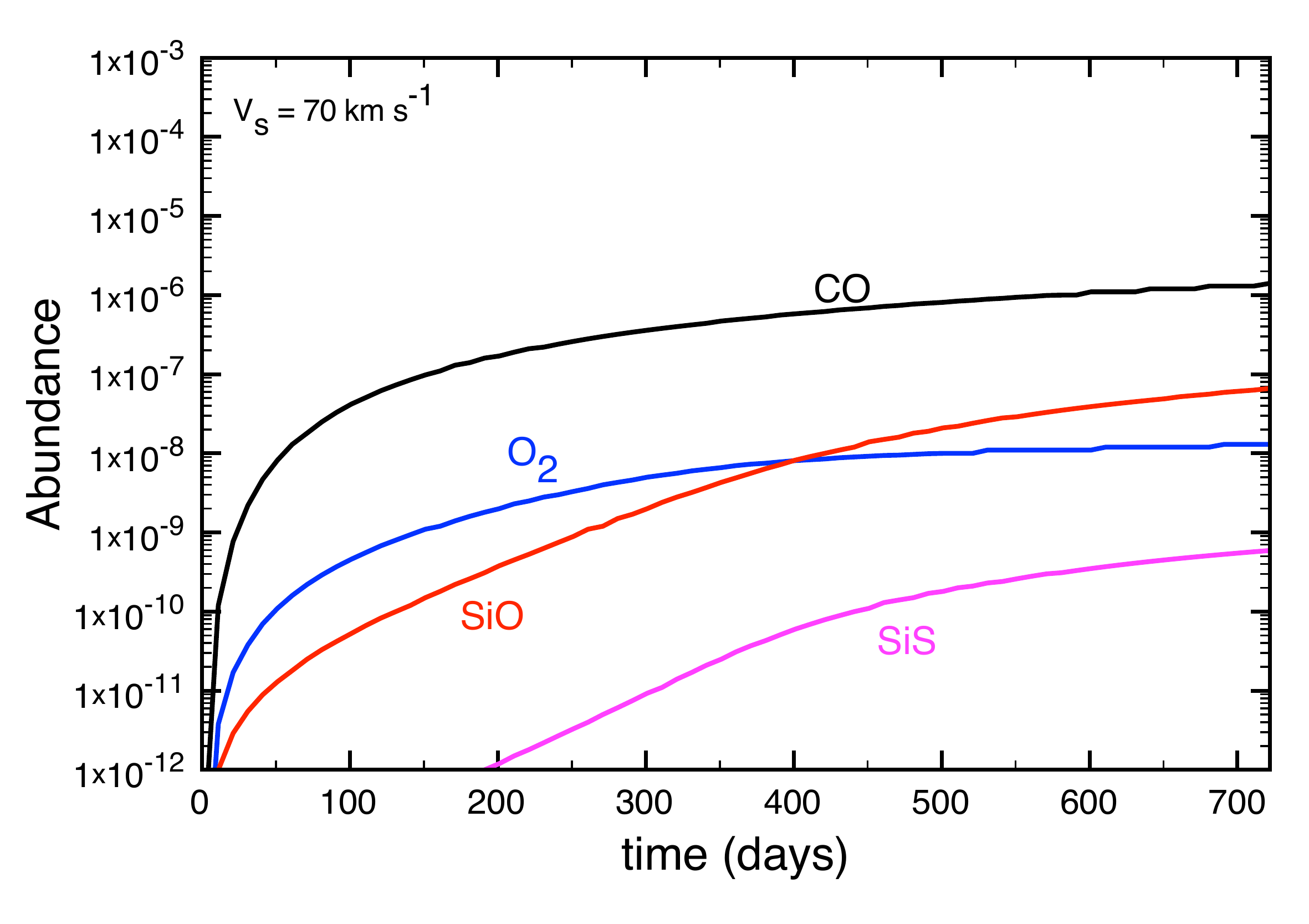}
\includegraphics[width=\columnwidth]{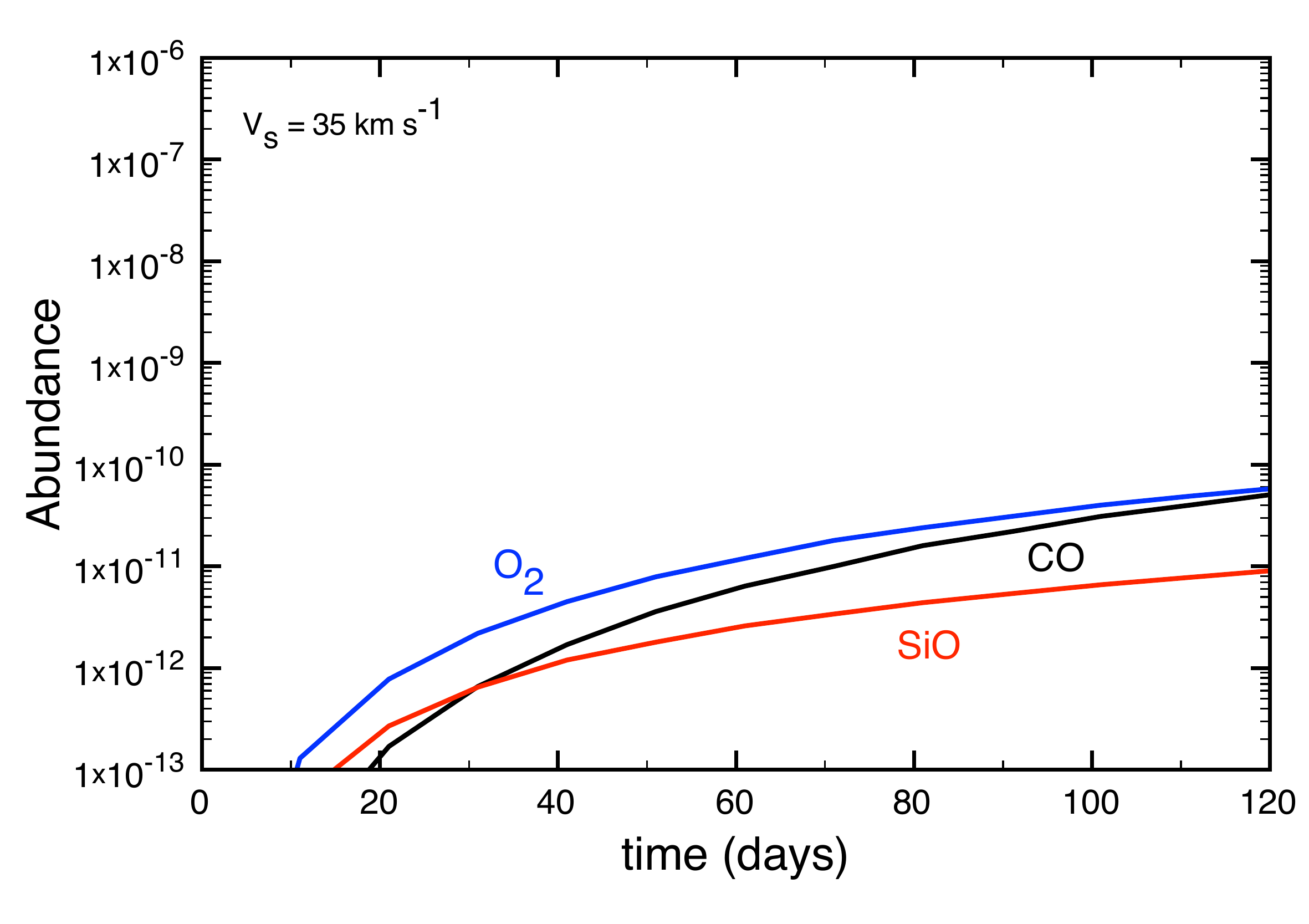}
  \caption{Abundances of molecules in the PIR (with respect to total gas number density) for various shock velocities. Top left: 140 \kms; Top right: 100 \kms; Bottom left: 70 \kms; Bottom right: 35 \kms. All shock models are from BS90. Molecules reform in the PIR from the gas phase but not at the level met for the pre-shock gas conditions. The abundances decrease with shock velocity because of the decreasing gas number density in the PIR and shorter PIR duration.  }
\label{fig6}
     \end{figure*}

FMKs have been observed in Cas A and interpreted as dense ejecta clumps that experience the crossing of the RS (\cite{chev78}). Their chemical composition includes a large fraction of oxygen atoms, along with neon, silicon, sulphur, argon and iron. This specific composition coincides with that of the oxygen-rich zones of the SN ejecta. Radiative shock theoretical models are able to reproduce optical and IR lines observed in these knots and provide information on the physical conditions at the reverse shock (\cite{ito81}, BS90, \cite{suth95}, DS10). We want to assess the impact on the RS on a FMK in terms of molecule and dust cluster reprocessing, and thus consider an oxygen-rich clump whose chemical composition is derived from our SN ejecta study presented in \S \ref{highden}. We follow the chemistry in the post-shock gas using the analytical model for the reverse shock illustrated in Figure \ref{fig1} and described in \S \ref{revsh}. We consider the five shock velocities of Table \ref{tab5}. According to BS90, oxygen atoms are fully ionised to different ionisation states in the HR. The high HR temperature (\env\ 10$^6$ K) ensures that all molecule and dust clusters present in the pre-shock clump are destroyed by the RS and atoms are ionised in the HR. We then model the chemistry in the PIR, assuming all atoms are singly ionised as initial conditions and study the chemistry, including the potential reformation of small dust clusters.

Abundances for the prevalent atoms, ions, and molecules in the PIR are shown in Figure \ref{fig5} for the 200 \kms\ shock. This shock model corresponds to the PIR conditions derived by DS10 with a different initial chemical composition. The ionisation of O atoms by the UV field coming from the HR is effective at maintaining a high ionisation fraction until \env\ 150 days, despite the rapid recombination to O atoms before \env\ 25 days. O$^+$ ions then recombine and exchange charge with other atoms to sustain an ionisation fraction of \env\ $0.03$, specifically through charge exchange reaction with Mg. Molecules chiefly reform from the gas phase in the PIR and CO has the largest abundance, followed by SiO, O$_2$, and SiS. The formation processes involved are radiative association reactions such as Reactions \ref{R1}, \ref{R2} or \ref{R4}. Other bimolecular processes that are active in the dense ejecta of Type II-P SNe are not efficient at the low gas number density of the PIR (in this case $n_{gas} = 10^6$ \cmc), e.g., Reaction \ref{R3}. CO, O$_2$ and SiS reform with lower abundances by a factor $\ge 1000$ than their initial values in the clump before the passage of the RS. SiO is an exception because the reformed gas-phase abundance is higher than its pre-shock value. As seen in \S \ref{highden}, SiO is quickly depleted into dust clusters in the SN ejecta that are later destroyed by the RS. The PIR conditions are then favourable to the formation of SiO from the gas phase, which traces the destruction of silicate clusters by the RS.

Similar trends operate for our lower shock velocity models shown in Figure \ref{fig6}. A lower shock speed combined with a shorter PIR duration result in the reformation of molecules with lower abundances as the RS velocity decreases. The physical conditions pertaining to the 140 \kms\ shock of BS90 in Figure \ref{fig6} are similar to the PIR model derived for the 200 \kms\ by DS10, and the species abundances for both models are akin (see Figure \ref{fig5}). CO is by far the more abundant species that reforms from the gas phase via Reaction \ref{R2}. The faster the RS, the larger the molecular content after the shock passage and the greater the chemical complexity of the PIR. When we consider the high PIR temperature derived by BS90 for their shock models with conduction ($T_2=4500$ K), no major changes in the abundance trends occur because most of the chemical routes that control the formation of molecules at the low gas number densities of the PIR are radiative association reactions. These processes are slow and temperature-independent. 

\begin{figure}
   \centering
   \includegraphics[width=\columnwidth]{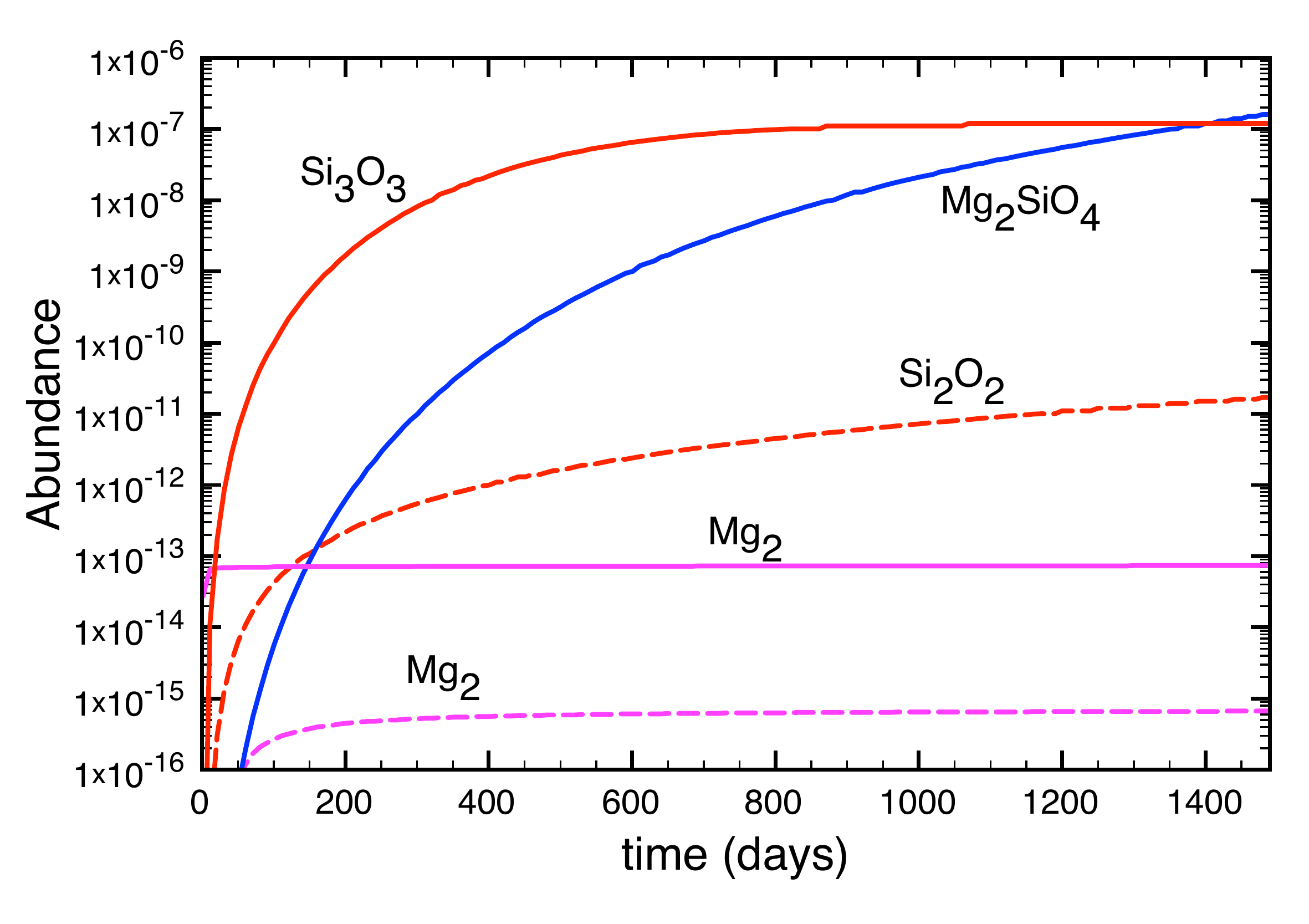}
      \caption{Abundances of dust clusters in the PIR (with respect to total gas number density) for the 140 \kms\ reverse shock. Dashed lines and full lines are for a gas density of $10^7$ \cmc\ and $10^9$ \cmc, respectively. Even when density enhancements are considered, no dust clusters reform in significant amounts.  }
 \label{fig7}
   \end{figure}

Dust clusters reform in the PIR from the gas phase for all shock velocities but with extremely low abundances ($\le 10^{-20}$) because of the low gas density involved. We explore the possibility of synthesising dust clusters by artificially increasing the gas number density in the PIR for the 140 \kms\ RS. We consider two PIR enhanced densities, $10^7$ \cmc\ and $10^9$ \cmc, and results are presented in Figure \ref{fig7}. We see that even the largest gas density fails at building up significant amounts of dust clusters, with a maximum abundance of \env\ $10^{-7}$ for forsterite and alumina dimers and (SiO) trimers. The PIR corresponds to a region where both the gas temperature and number density are high compared with values characterising the pre-shock gas in the clump and non-shocked gas in the remnant. This environment is thus more conducive to dust cluster formation from the gas phase. If dust clusters are unable to form from the gas phase in this region, we can then anticipate that the dust grains destroyed by the RS crossing dense gas clumps are unable to reform in the remnant from the gas phase.  

\begin{figure}
   \centering
   \includegraphics[width=\columnwidth]{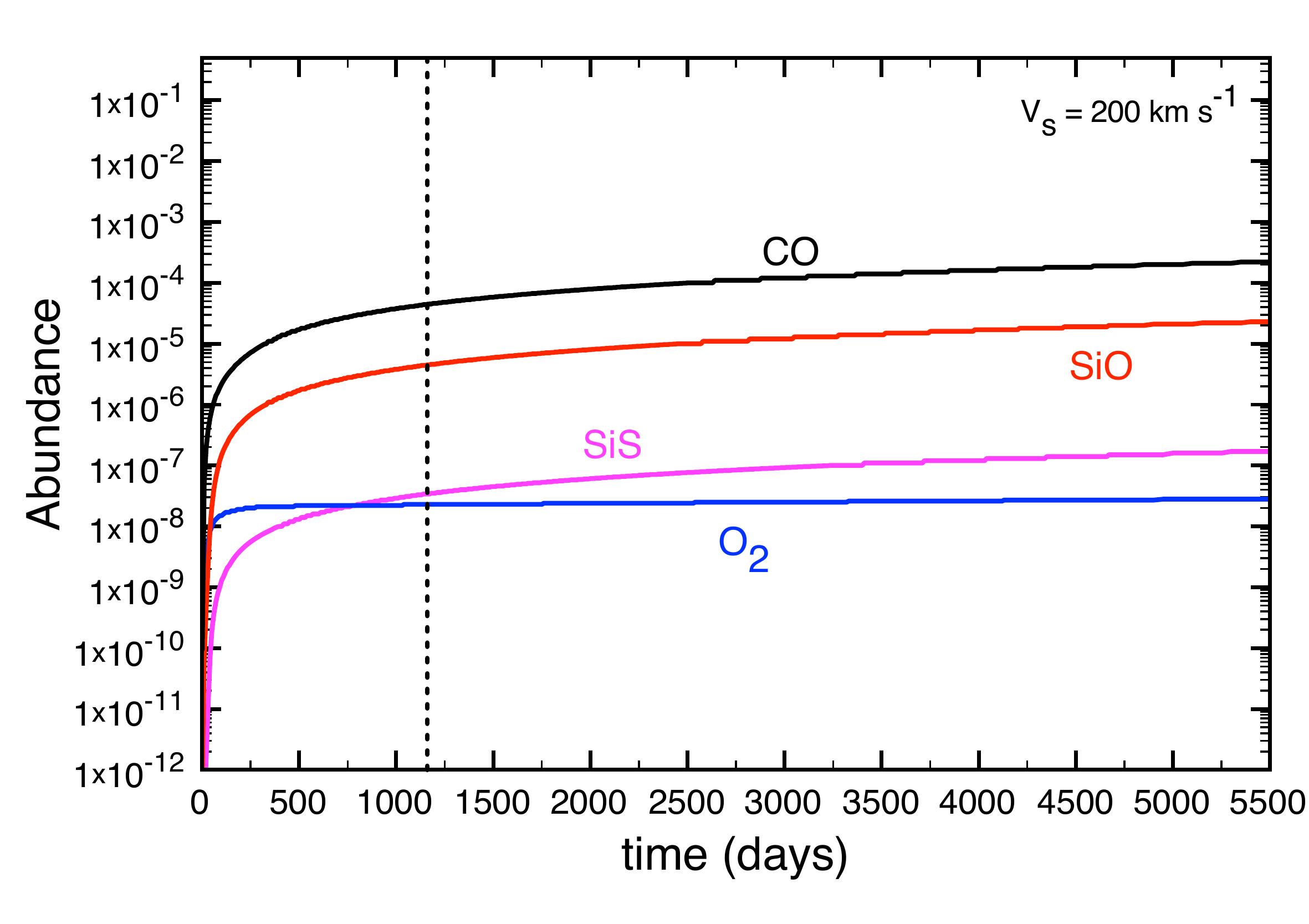}
      \caption{Abundances of molecules (with respect to total gas number density) as a function of time in the post-shock region of the 200 \kms\ RS. The vertical dotted line indicates $t = t_{PIR}$, and results for $t < t_{PIR}$ are similar to those of Fig. \ref{fig5}. }
 \label{fig8}
   \end{figure}

We have so far considered the chemistry of the PIR region after the passage of the RS. Since molecules reform but dust clusters do not for the standard physical conditions of the PIR region given by Table \ref{tab5}, we want to investigate whether molecules could keep forming over a time span $t >> t_{PIR}$. Following Sutherland \& Dopita (1995), the clump-crushing time is given by $\tau_{cc} = {\chi}^{1/2}/v_0$, where $v_0$ is the pre-RS velocity and $\chi$ the density contrast defined in \S~\ref{revsh}. According to Silvia et al. (2010), typical clump-crushing times for the $\chi$ factors and pre-RS velocities considered in this study range from 6 to 100 years. We thus consider the RS shock model with $V_s=200$ \kms, keep the PIR number density constant over a period of \env\ 15 years, and decrease the gas temperature to 300 K to account for radiative cooling. The abundances of the prevalent molecules are shown in Figure \ref{fig8}. Most molecules slowly keep forming from radiative association processes while dust clusters form with negligible abundances. The final abundance values for molecules at $\tau_{cc}$ remain all below $10^{-3}$ for $6 < \tau_{cc}< 100$~years. We can thus conclude that molecules quickly reform in the dense and warm PIR region and continue forming in the post-RS gas till $t = \tau_{cc}$. We also confirm that dust clusters are unable to form from the gas phase in the PIR and at later times in the shocked clump. 

\section{Discussion}
\label{dis}
We have investigated the synthesis of molecules and dust clusters in the stratified ejecta of the Cas A Type IIb SN progenitor. Focussing on the oxygen-rich material formed in the oxygen-rich zones of the ejecta, we have studied their processing by the reverse shock in a dense clump of the Cas A remnant. Our results highlight the following points

\begin{itemize} 
\item The stratified, homogeneous ejecta of a Type IIb SN is too diffuse to form large amounts of molecules and dust. For normal ejecta conditions, only \env\ $10^{-2}$~\Ms\ of molecules and \env\ $10^{-5}$~\Ms\ of silicon carbide and silica dust clusters form. When the gas density is raised, the masses of molecules and clusters also increase, and most importantly, the chemical complexity of the dust composition is enhanced. \item In the remnant, molecules and dust clusters in the gas phase of oxygen-rich clumps are destroyed by the crossing of the reverse shock but chemical species, including CO, SiO, SiS, and O$_2$ reform in the post-shock gas. Stronger shocks lead to denser post-shock photo-ionised regions and thus higher abundances of reformed molecules.  
\item No clusters of silicate, silica, silicon carbide, amorphous carbon, or pure metal grains can reform in the back of the shock because the post-shock gas densities are too low to permit the chemical build-up of dust clusters and the formation of dust grains. 
\end{itemize}
The main results of this study point to the importance of density in the formation processes of molecules and dust in SNe and their remnants. In the homogeneous ejecta of a Type IIb SN, we find that no molecular cluster precursors to dust could form at the low gas density involved. This points to the complexity of the dust formation processes at a microscopic scale, and the fact that dust nucleation is driven by non-equilibrium chemistry in the gas-phase. Existing studies based on the CNT formalism ignore the dust nucleation bottleneck and indicate fairly large masses of grains formed in Type IIb SNe (e.g., \cite{noz10}). Our present results are in contrast with these studies and show that the formation of dust is extremely gas density-dependent. This dependency does not pertain to the condensation phase of dust clusters but to their formation phase, i.e., nucleation out of the gas-phase. Increase in gas density by a factor of $200-2000$ permits an efficient nucleation of dust grains and indicates the need for non-homogenous, Type IIb SN ejecta in the form of clumps or knots. 

The chemical complexity of the dust grains that form in SN ejecta is also density-dependent and grows with increasing gas density. In particular, carbon dust only forms at high gas number densities and traces the densest SN ejecta conditions. In Cas A, the IR spectra measured by Spitzer indicate a variety of spectral signatures ascribed to warm dust which vary according to position and observed atomic lines in the remnant (\cite{rho08, ar14}). These grains include silicates, alumina, pure metals, carbides, iron sulphide, and amorphous carbon and have been processed by the RS, but they originate from the SN ejecta that led to Cas A. According to our results, the SN ejecta gas must be in the form of dense clumps, with density contrast of at least 200 compared to the homogeneous Type IIb SN ejecta, to properly account for the dust chemical diversity in Cas A inferred from the IR data. More generally, the chemical type of dust detected in SNe and SNRs can be used as a tracer of the physical conditions of the dust birthplace, a denser gas leading to dust grains of greater chemical variety.   

Recent observations with Herschel of high excitation energy rotational lines of CO in one region of Cas A where CO was already detected at the RS position with Spitzer have brought evidence of hot and dense gas in the remnant (\cite{wal13}). According to the line analysis, this O/CO-rich region is characterised by warm  and dense gas ($n_{gas} \sim 10^6$ \cmc), and probably corresponds to post-RS gas. Such conditions are found in the wake of our 200 and 140 \kms\  RS crossing a dense clump where CO reforms on a time scale of \env\ 100 days. These observations therefore support the present model results which indicate that molecules, including CO and SiO, reform after the RS passage in sufficient amount to warrant detection. The molecular content of Cas A should then include two components reflecting various evolutionary phases of the remnant. A cold component would comprise species formed in the SN clumpy ejecta (e.g., SiS, CO, O$_2$, and SO), that have not yet been processed by the RS and have large abundances. Such a cool component has been confirmed in SN1987A by the detection of low excitation rotational lines of CO with ALMA (\cite{kam13}). A warm/hot component would coincide with shocked clumps and newly reformed molecules (e.g., CO, SiO) in the post-reverse shock gas, as seen in Cas A with Herschel. The warm CO component should have a lower abundance than its cold counterpart. The amount of reformed molecules depends on the shock strength, the faster shocks producing larger quantities of chemical species in the post-shock gas. For all shock speeds, dust clusters are unable to reform in sufficient quantities in the post-shock gas to guarantee a viable dust formation pathway from the gas phase in SN remnants. 

\begin{acknowledgement}
The authors thank the anonymous referee for helpful comments. C.B. thanks the support from the Swiss National Science Foundation through the subside 20GN21$-$128950 linked to the CoDustMas network of the European Science Foundation Eurogenesis programme. 
\end{acknowledgement}

\clearpage

\end{document}